\documentclass[12pt]{article}
\usepackage{subfigure}
\usepackage{graphicx}
\usepackage{harvard}
\usepackage{amsfonts}
\usepackage{amsmath}
\usepackage{amssymb}
\usepackage{color}
\usepackage{verbatim}

\def\Bmp#1{ \begin{minipage}{#1} }
\def\Bmpc#1{ \begin{minipage}[c]{#1} }
\def\Bmpt#1{ \begin{minipage}[t]{#1} }
\def\Bmpb#1{ \begin{minipage}[b]{#1} }
\def\Emp{ \end{minipage} }

\def\E{{\mathcal{E}}}

\def\P{{\mathcal{P}}}
\def\R{{\mathcal{R}}}

\def\K{{\mathcal{K}}}
\def\S{{\mathcal{S}}}

\def\tpKP{\tilde{\psi}_{\K_0,\P_0}}

\def\tf0{\tilde{\varphi}_{0}}

\def\RR{{\mathbb{R}}}

\def\x{{\bf x}}

\def\u{{\bf u}}
\def\0{{\bf 0}}

\def\bnabla{\mathbf{\nabla}}

\def\Dpartial#1#2{ {\frac{\partial #1}{\partial #2} }}

\begin{document}
\title{Vortices, Maximum Growth and the Problem of Finite-Time Singularity
  Formation}

\author{Diego Ayala and Bartosz Protas\thanks{Email address for correspondence: bprotas@mcmaster.ca}
\\
Department of Mathematics and Statistics, McMaster University \\
Hamilton, Ontario, L8S 4K1, Canada
}
\maketitle

\begin{abstract}
  In this work we are interested in extreme vortex states leading to
  the maximum possible growth of palinstrophy in 2D viscous
  incompressible flows on periodic domains. This study is a part of a
  broader research effort motivated by the question about the
  finite-time singularity formation in the 3D Navier-Stokes system and
  aims at a systematic identification of the most singular flow
  behaviors. We extend the results reported in \citeasnoun{ap13a}
  where extreme vortex states were found leading to the growth of
  palinstrophy, both instantaneously and in finite-time, which
  saturates the estimates obtained with rigorous methods of
  mathematical analysis. Here we uncover the vortex dynamics
  mechanisms responsible for such extreme behavior in time-dependent
  2D flows. While the maximum palinstrophy growth is achieved at short
  times, the corresponding long-time evolution is characterized by
  some nontrivial features, such as vortex scattering events.
\end{abstract}

\vspace{2pc}
variational optimization


\section{Introduction}
\label{sec:intro}

The research program referred to in this study is motivated by the
questions concerning the existence of smooth solutions to the 3D
Navier-Stokes system for arbitrarily large times, corresponding to
smooth initial data of arbitrary size. To date, smooth solutions have
been proved to exist for {\em finite} times only \cite{d09}, leaving
open the possibility of a spontaneous formation of singularities in
finite time. We mean by this the loss of regularity of the solution
manifested by the ``blow-up'' of its certain norms. The importance of
this problem has been recognized by the Clay Mathematics Institute
which identified it as one of the ``millennium problems'' \cite{f00}.
There are also analogous questions concerning the existence of smooth
solutions to the 3D Euler equation (\citeasnoun{bt07},
\citeasnoun{g08a}), as well as the equations describing
quasi-geostrophic flows (\citeasnoun{oy97}, \citeasnoun{cfmp05},
\citeasnoun{s11}) and magnetohydrodynamic phenomena \cite{cm00}.

It is believed that, should such blow-up occur in finite time, it
should be associated with the formation and amplification of
small-scale vortex structures. Indeed, a number of different vortex
states have been proposed as the initial data for both the Euler
system (e.g., \citeasnoun{bmonmu03}, \citeasnoun{ps90},
\citeasnoun{k93}, \citeasnoun{p01}, \citeasnoun{gg12},
\citeasnoun{opc12}) and the Navier-Stokes system (e.g.,
\citeasnoun{bmonmu03}, \citeasnoun{b91}, \citeasnoun{opc12}) which
might possibly lead to singularities in finite time, although the
computational evidence is in either case not conclusive \cite{g08a}.
As regards probing the flow behavior, in addition to tracking the time
evolution of various vorticity norms, some attention has also been
given to geometric criteria (\citeasnoun{h09}, \citeasnoun{gg12}) and
to characterization of the solutions in the complex plane through the
width of the analyticity strip (\citeasnoun{p10}, \citeasnoun{bb12}).
All of the aforementioned candidate flows were postulated in a rather
ad-hoc fashion based on purely physical arguments. The goal of the
research program this study is a part of is to perform search for such
extreme vortex structures in a systematic manner using variational
methods of mathematical optimization. It is this aspect of bridging
the mathematical theory with large-scale computations that
distinguishes the present research program, initiated by
\citeasnoun{ld08}, from earlier efforts.

The key observation is that the question about finite-time singularity
formation can be rephrased in terms of boundedness of certain norms of
the solution. More precisely, if $\u(t,\cdot)$, $t>0$, is a 3D
velocity vector field corresponding to some smooth initial data
$\u_0(\cdot) = \u(0,\cdot)$, then it is well-known that the boundedness of
the enstrophy $\E(t) := \frac{1}{2}\int_{\Omega} | \bnabla \times
\u(t,\x)|^2 \, d\Omega$ will guarantee smoothness of the solution up
to time $t$ \cite{ld08}.  Using methods of functional analysis, the
rate of growth of enstrophy can be estimated as
\begin{equation}
\frac{d\E(t)}{dt} < C \E(t)^3, \qquad \textrm{where} \ C > 0,
\label{eq:dEdt}
\end{equation} 
which, upon
integration with respect to time, leads to
\begin{equation}
\E(t) \le  \frac{\E(0)}{\sqrt{1 - 4 t \frac{C \E(0)^2}{\nu^3}}}.
\label{eq:Et}
\end{equation}
For brevity, $C$ will hereafter denote a generic positive constant
which may assume different numerical values in different instances.
We note that upper bound \eqref{eq:Et}, which is the sharpest result of
this kind available to date \cite{d09}, blows up in finite time $t^* =
\nu^3 / (4 C \E(0)^2)$, and, based on this estimate alone, singularity
formation cannot be ruled out. The question about the possibility of
finite-time blow-up can be thus cast in terms of sharpness of estimate
\eqref{eq:Et} (by ``sharpness'' we mean existence of solutions which
saturate a given inequality bound). Such problems can be studied
using variational optimization methods allowing one to find the vortex
structures which are the most singular in a suitable sense.
\citeasnoun{ld08} used this approach to demonstrate that instantaneous
estimate \eqref{eq:dEdt} is in fact sharp (up to a numerical
prefactor).  In order to assess sharpness of the finite-time estimate
\eqref{eq:Et}, a question which may hold valuable insights concerning
the blow-up problem, one would need to solve the following
optimization problem
\begin{equation}
\max_{\u_0 \in H^1(\Omega)} \E(T) \ \ \textrm{subject to} \ \ \E(0) = \E_0, 
\label{eq:maxE}
\end{equation} 
where $\E_0>0$, $T>0$ are given, $H^1(\Omega)$ is the Sobolev space of
functions with square-integrable gradients \cite{af05} and the control
variable is the initial data $\u_0$. While numerical solution of
optimization problem \eqref{eq:maxE} for a broad range of $\E_0$ and
$T$ is a formidable computational task, it appears within reach of
modern methods of PDE-constrained optimization and remains the
ultimate long-term goal of this research program. Such problems are
typically solved using gradient-based methods where the gradient
information is determined based on the solution of a suitable adjoint
system.

Analogous questions can in fact be also formulated in the case of
simpler systems, such as the 1D Burgers and 2D Navier-Stokes
equations. It is known that both 1D Burgers and 2D Navier-Stokes
systems have solutions which remain smooth for arbitrary times
\cite{kl04}, hence there is no question about their finite-time
blow-up. However, for such systems there also exist analytical upper
bounds on the growth of various quantities, analogous to
\eqref{eq:dEdt} and \eqref{eq:Et}, and since they are obtained with
similar methods as for the 3D Navier-Stokes system, the questions
about their sharpness are in fact quite relevant. Needless to say, the
computational tasks arising in the solution of such variational
optimization problems in 1D and 2D are much more tractable than in the
3D setting. Hence, from the fundamental perspective, there is a
significant interest in studying how sharp the mathematical analysis
is in describing the extreme behavior of 1D and 2D flows. While the
main interest, especially from the point of view of the singularity
formation, is to assess the sharpness of the {\em finite-time}
estimates, they are obtained from the instantaneous estimates which
provide upper bounds on the rate of growth of a given quantity at a
{\em fixed} instant of time. We add that, in contrast to 1D and 3D
flows, in 2D flows on unbounded or doubly-periodic domains the
enstrophy can only decrease and the relevant quadratic quantity is the
palinstrophy (which will be formally defined below). The best
estimates available for the different problems are summarized in Table
\ref{tab:estimates} in which we indicate whether or not they have been
found to be sharp in earlier investigations and also highlight
outstanding open questions.  Details concerning the derivation of the
different estimates can be found in \citeasnoun{ld08},
\citeasnoun{ap11a} and \citeasnoun{ap13a}.

\begin{table}
\begin{center}
\hspace*{-1.1cm}
\begin{tabular}{l|c|c}      
  &  \Bmp{3.0cm} \small \begin{center} {\sc Best Estimate} \\ \smallskip \end{center} \Emp   
  & \Bmp{3.8cm} \small \begin{center} {\sc Sharpness}  \end{center} \Emp \\  
  \hline
  \Bmp{2.5cm}  \small {\begin{center} \smallskip 1D Burgers  \\ instantaneous \smallskip \end{center}} \Emp &  
  \small {$\frac{d\E}{dt} \leq \frac{3}{2}\left(\frac{1}{\pi^2\nu}\right)^{1/3}\E^{5/3}$}  & 
  \Bmp{3.8cm} \small {\begin{center} \smallskip {\sc Yes} \\ \cite{ld08}  \smallskip  \end{center}} \Emp \\ 
  \hline 
  \Bmp{3.0cm} \small {\begin{center} \smallskip 1D Burgers  \\ finite-time \smallskip \end{center}} \Emp &  
  \small {$\max_{t \in [0,T]} \E(t) \leq \left[\E_0^{1/3} + \frac{1}{16}\left(\frac{1}{\pi^2 \nu}\right)^{4/3}\E_0\right]^{3}$} &  
  \Bmp{3.9cm} \small {\begin{center} \smallskip {\sc No} \\ \cite{ap11a} \smallskip  \end{center}} \Emp \\ 
  \hline 
  \Bmp{3.0cm} \small {\begin{center} \smallskip 2D Navier-Stokes  \\ instantaneous \smallskip\end{center}} \Emp &  
  \Bmp{7.0cm} \smallskip \centering \small $\frac{d\P(t)}{dt} \le \frac{C_2}{\nu} \K^{1/2}\P^{3/2}$ \smallskip \Emp& 
  \Bmp{3.8cm} \small {\begin{center} \smallskip {\sc Yes} \\ \cite{ap13a} \end{center}} \Emp  \\ 
  \hline 
  \Bmp{3.0cm} \small \begin{center} \smallskip 2D Navier-Stokes \\ finite-time \smallskip\end{center} \Emp &  
  \Bmp{7.0cm} \smallskip \centering \small  $\max_{t>0} \P(t) \le \left[\P_0^{1/2} + \frac{C_2}{4\nu^2}\K_0^{1/2}\E_0\right]^2$ \smallskip \Emp  & \Bmp{3.8cm} \small {\begin{center} \smallskip {\sc Yes} \\ \cite{ap13a} \end{center}} \Emp  \\ 
  \hline 
  \Bmp{3.0cm} \small {\begin{center} \smallskip 3D Navier-Stokes  \\ instantaneous  \smallskip \end{center}} \Emp &  
  \small {$\frac{d\E(t)}{dt} \le \frac{27 C^2}{32 \nu^3} \E(t)^3$} & \Bmp{3.8cm} \small {\begin{center} \smallskip {\sc Yes} \\ \cite{ld08} \smallskip  \end{center}} \Emp  \\ 
  \hline 
  \Bmp{3.0cm} \small \begin{center} \smallskip\smallskip 3D Navier-Stokes  \\ finite-time \smallskip \end{center} \Emp &  
  \small $\E(t) \le \frac{\E(0)}{\sqrt{1 - 4 \frac{C \E(0)^2}{\nu^3} t}}$ & \Bmp{3.5cm} \centering {???} \smallskip\smallskip \Emp
\end{tabular}
\end{center}
\caption{ 
  Summary of the best estimates available to date for the 
  instantaneous rate of growth and the growth over finite time of 
  enstrophy and palinstrophy in 1D Burgers, 2D and 3D Navier-Stokes 
  systems (we refer the reader to 
  \protect\citeasnoun{ap13a} 
  for a more complete version of this table featuring some additional estimates in 2D).
}
\label{tab:estimates}
\end{table}

In the present study we offer a vortex dynamics perspective on the
recent results of \citeasnoun{ap13a} where it was shown that one of
the estimates for the instantaneous rate of growth $d\P/dt$ of the
palinstrophy in 2D is in fact saturated by suitably constrained
families of vorticity fields. Moreover, in that study it was also
demonstrated that when these maximizing vortex states are used as the
initial data for the 2D Navier-Stokes system, then the ensuing flow
evolution actually saturates the corresponding finite-time estimate.
This is an intriguing finding which is at odds with what was observed
in \citeasnoun{ap11a} in the case of the 1D Burgers equation, and may
also be important for the ultimate question concerning the sharpness
of 3D finite-time estimate \eqref{eq:Et}. In this investigation we
analyze in some detail the structure of the optimal vortex states
responsible for such extreme events. The structure of the paper is as
follows: in Section \ref{sec:problem} we formally state the problem;
in the following Section we briefly recall some of the relevant
results from \citeasnoun{ap13a} concerning the vorticity fields
maximizing $d\P/dt$ under different constraints; in Section
\ref{sec:ft} we analyze the corresponding time evolution for both
short and long times; finally, discussion and conclusions are deferred
to Section \ref{sec:final}.

\section{Two-Dimensional Navier-Stokes System and the Palinstrophy
  Growth}
\label{sec:problem}

We are concerned with the motion of a viscous incompressible fluid on
a 2D periodic domain $\Omega = [0,1]\times[0,1]$ which is governed by
the Navier-Stokes system
\begin{subequations}
\label{eq:NS2D}
\begin{alignat}{2}
& \Dpartial{\omega}{t} + J(\omega,\psi) = \nu \Delta \omega   \qquad &
\textrm{in} \ (0,\infty) \times \Omega, \label{eq:NS2Da} \\
& - \Delta \psi = \omega \quad &
\textrm{in} \ (0,\infty) \times \Omega, \label{eq:NS2Db} \\
& \omega(0) = \omega_0 \quad &
\textrm{in} \ \Omega, \label{eq:NS2Dc}
\end{alignat}
\end{subequations}
where $\psi$ and $\omega$ are, respectively, the streamfunction and
(scalar) vorticity, whereas $\omega_0$ is the initial condition. In
system \eqref{eq:NS2Da}--\eqref{eq:NS2Dc} $\nu$ denotes the kinematic
viscosity (assumed fixed and equal to $10^{-3}$ in all the
  results presented below), $\Delta$ is the Laplacian operator and
$J(f,g) := \partial_x f \, \partial_y g - \partial_y f \, \partial_x
g$, defined for $f,g \; : \; \Omega \rightarrow \RR$, is the Jacobian
determinant. For simplicity, we will work with the streamfunction
$\psi(t,\cdot) \; : \; \Omega \rightarrow \RR$ as the state variable.
The following quadratic quantities will play a key role in our
analysis
\begin{eqnarray}
&\textrm{kinetic energy}  \qquad\quad
& \K(\psi(t)) = \frac{1}{2}\int_{\Omega} |\bnabla \psi(t,\x)|^2 \, d\Omega,
\label{eq:K} \\
&\textrm{enstrophy} 
& \E(\psi(t))  = \frac{1}{2}\int_{\Omega} (\Delta \psi(t,\x))^2 \, d\Omega,
\label{eq:E} \\
&\textrm{palinstrophy} 
&\P(\psi(t))  = \frac{1}{2}\int_{\Omega} |\bnabla \Delta \psi(t,\x)|^2 \, d\Omega.
\label{eq:P}
\end{eqnarray}
Noting that, for the evolution described by system
\eqref{eq:NS2Da}--\eqref{eq:NS2Dc}, we always have $d\E(t) / dt = - 2
\nu \P(t) < 0$, in 2D flows one is interested in the rate of growth of
the palinstrophy $\P$ which can be expressed using Navier-Stokes
equation \eqref{eq:NS2Da} as
\begin{equation}
\frac{d\P(t)}{dt} = \int_{\Omega} J(\Delta\psi,\psi) \Delta^2 \psi\, d\Omega
- \nu \, \int_{\Omega} (\Delta^2 \psi)^2 \, d\Omega =: \R_{\P}(\psi).
\label{eq:R}
\end{equation}
Using rigorous methods of mathematical analysis, this quantity can be
upper bounded as
\begin{equation}
\frac{d\P}{dt}  \le \frac{C}{\nu} \,\K^{\frac{1}{2}}\, \P^{\frac{3}{2}},
\label{eq:dPdt_KP}
\end{equation}
which leads to the following finite-time estimate
\begin{equation}
\max_{t > 0} \P(t) \, \le \, \left[\P^{1/2}_0 + \frac{C}{4\nu^2} \K^{1/2}_0 \E_0\right]^2
\label{eq:maxPt_Ayala}
\end{equation}
in which $\K_0 := \K(0)$, $\E_0 := \E(0)$ and $\P_0 := \P(0)$.  For
the derivation of these estimates and a discussion of other bounds on
$d\P/dt$ and $\max_{t>0}\P(t)$ the reader is referred to
\citeasnoun{ap13a}. Here we only remark that quantity $\R_{\P}(\psi)$,
cf.~\eqref{eq:R}, is intrinsically related to the stretching of the
vorticity gradients $\bnabla\omega$. As is well-known (see, e.g.,
\citeasnoun{pkb99}), the equation characterizing the evolution of
$\bnabla\omega$ features a quadratic stretching term reminiscent of
the ``vortex stretching'' term in the 3D vorticity equation.

\section{Vortex States Maximizing the Instantaneous 
Rate of Growth of Palinstrophy}
\label{sec:maxdpdt}

In order to provide the context for the main results of this paper
presented in the next Section, we briefly recall here some of the most
important findings from \citeasnoun{ap13a} concerning the vortex
states maximizing $d\P/dt$, cf.~\eqref{eq:R}. The key question is
whether estimate \eqref{eq:dPdt_KP} can be saturated by vortex states
with prescribed energy $\K_0$ and palinstrophy $\P_0$. To address this
question, such maximizing vorticity fields $-\Delta\tpKP$ were sought
via solution of the following constrained optimization problem
\begin{eqnarray}
\tilde{\psi}_{\K_0,\P_0} & = \mathop{\arg\max}_{\psi\in\S_{\K_0,\P_0}} \,\R_{\P_0}(\psi)  \label{eq:optR_KP} \\ 
\S_{\K_0,\P_0} & = \left\{\psi \in H^4(\Omega) : \frac{1}{2}\int_\Omega|\bnabla\psi|^2\,d\Omega = \K_0, \ \frac{1}{2}\int_\Omega|\bnabla\Delta\psi|^2\,d\Omega = \P_0 \right\}  \nonumber
\end{eqnarray}
(for some technical reasons, maximization is performed in the Sobolev
space $H^4(\Omega)$ of functions with square-integrable derivatives of
order up to 4 \cite{af05}). We note that the main difficulty in solving problem
\eqref{eq:optR_KP} is the presence of two nonlinear constraints which
means that the maximizers $\tpKP$ need to be found at the intersection
of two nonlinear constraint manifolds. We also add that the values of
the constraints are linked through a nested Poincar\'e's inequality
$\K_0 \le (2\pi)^{-4} \, \P_0$.

\begin{figure}
\setcounter{subfigure}{0}
\begin{center}
\subfigure[]{\includegraphics[width=0.45\textwidth]{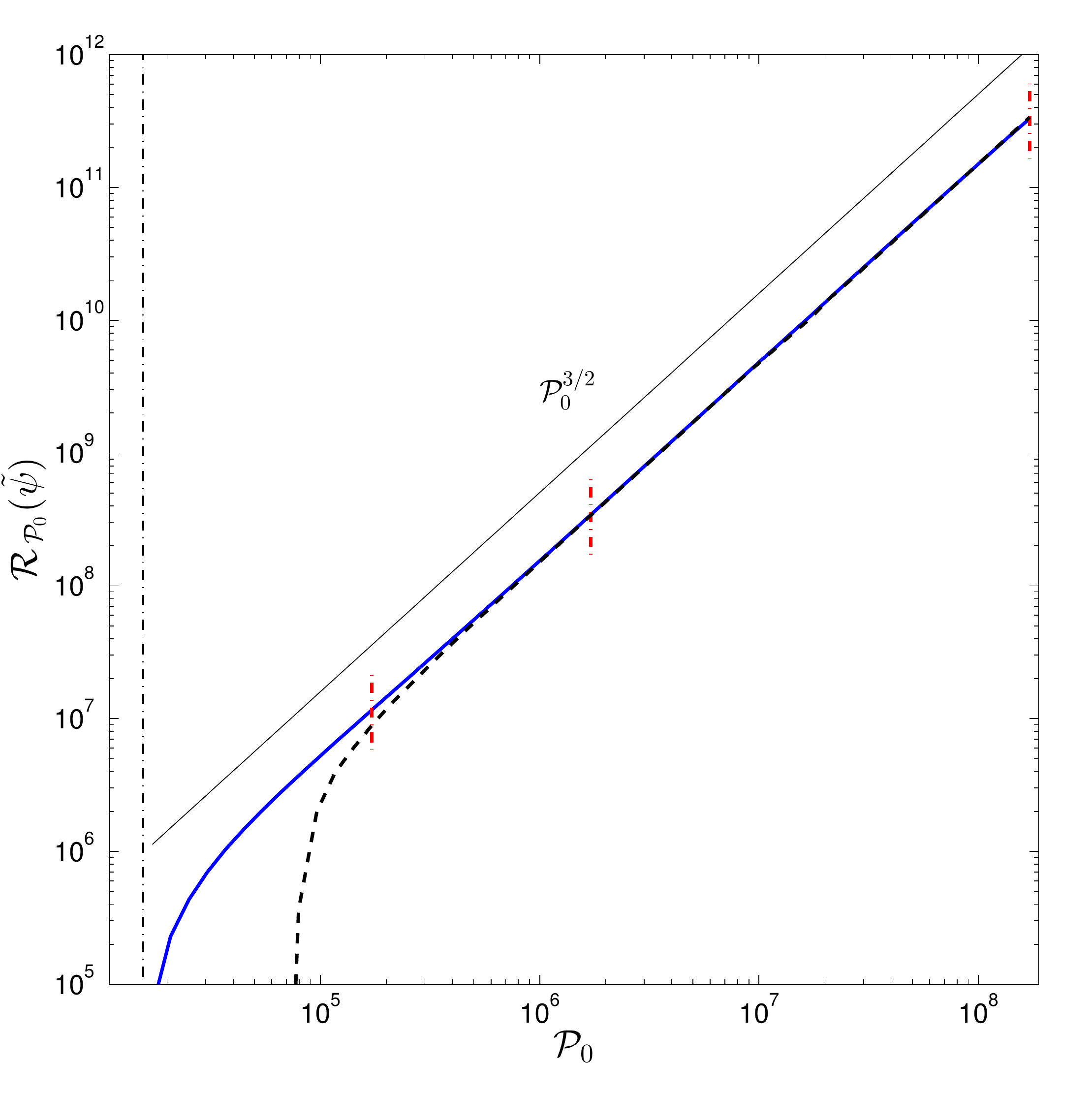}}
\subfigure[]{\includegraphics[width=0.45\textwidth]{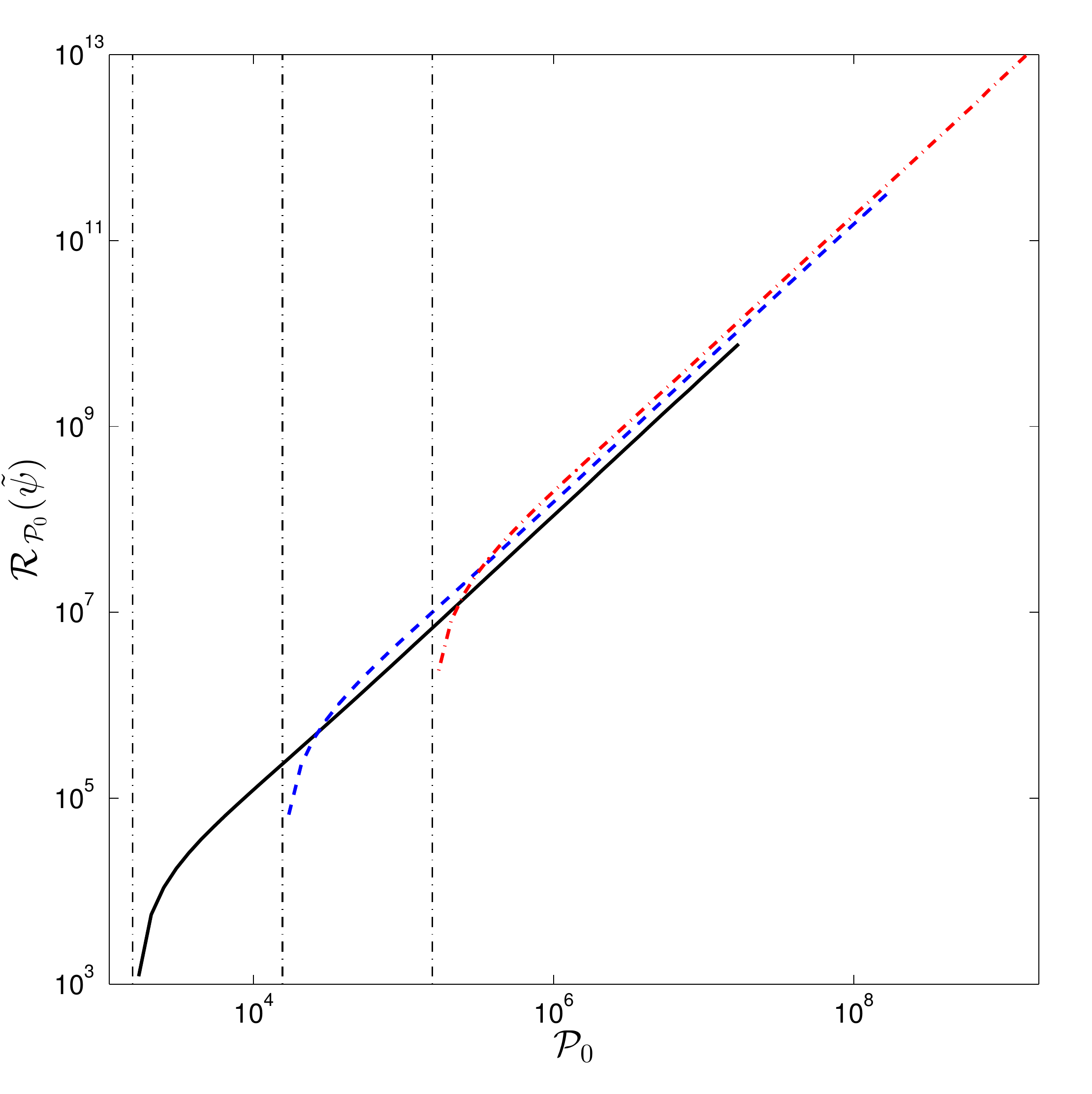}}\\
\subfigure[]{\includegraphics[width=0.3\textwidth]{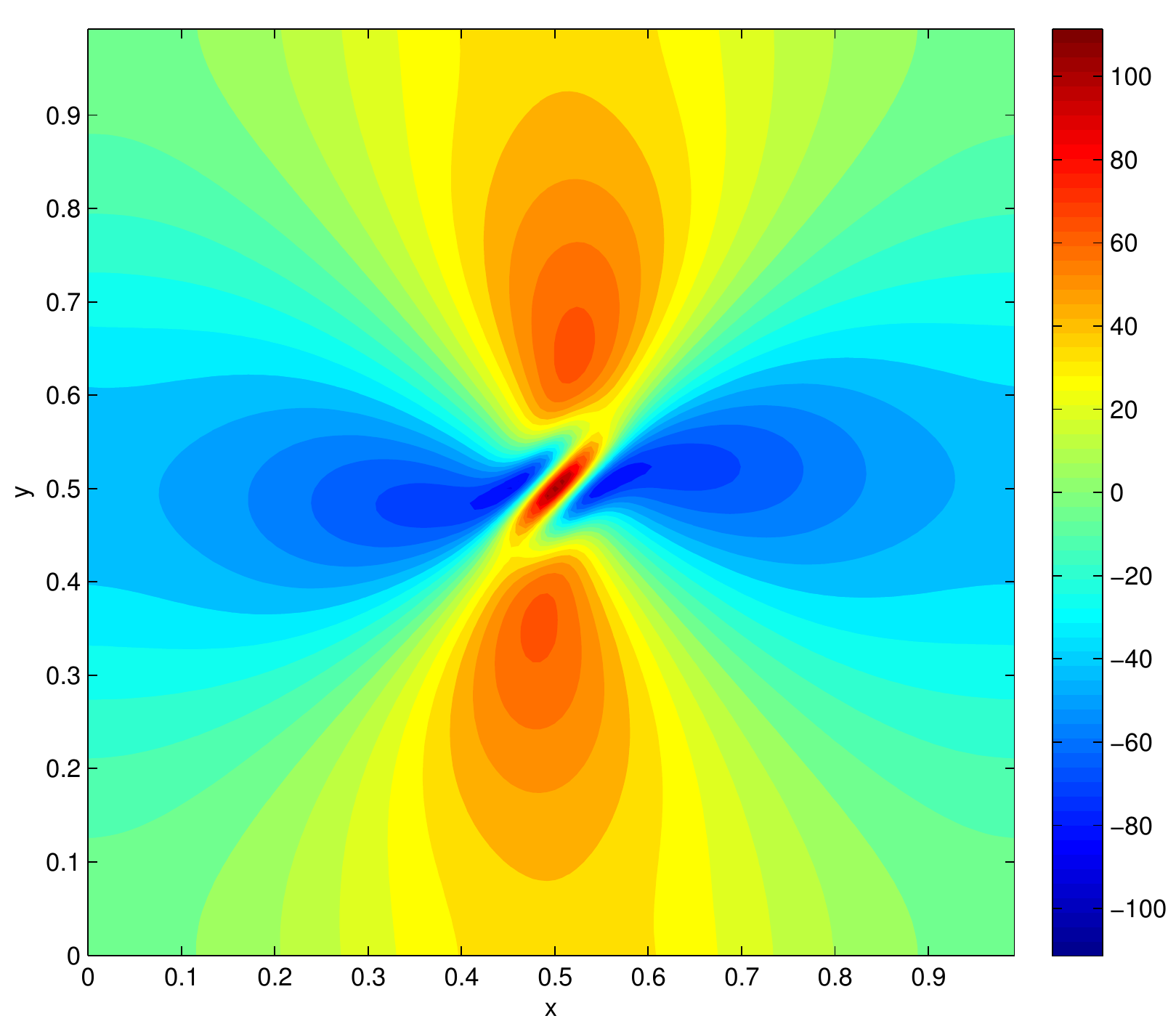}}
\subfigure[]{\includegraphics[width=0.3\textwidth]{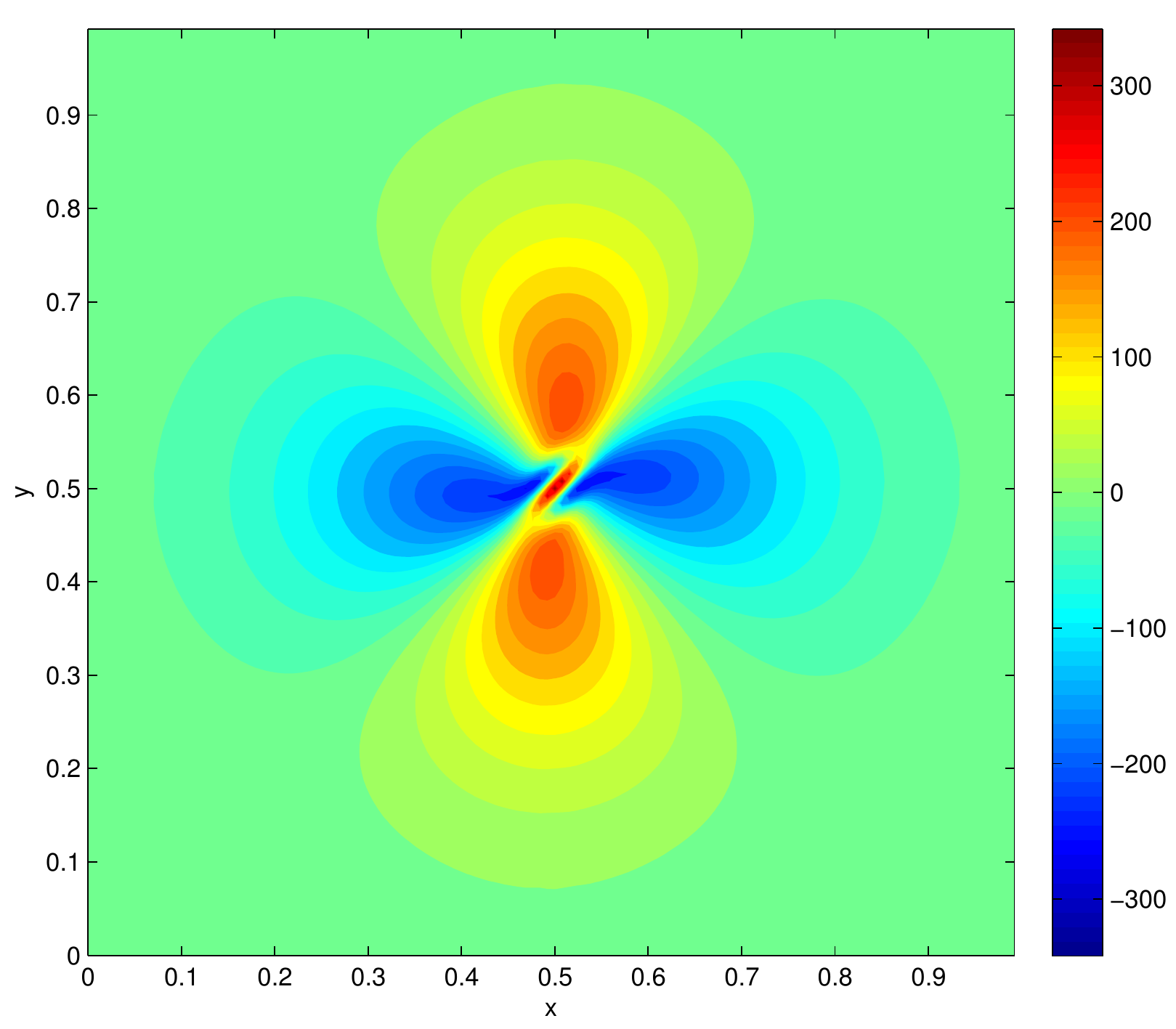}}
\subfigure[]{\includegraphics[width=0.3\textwidth]{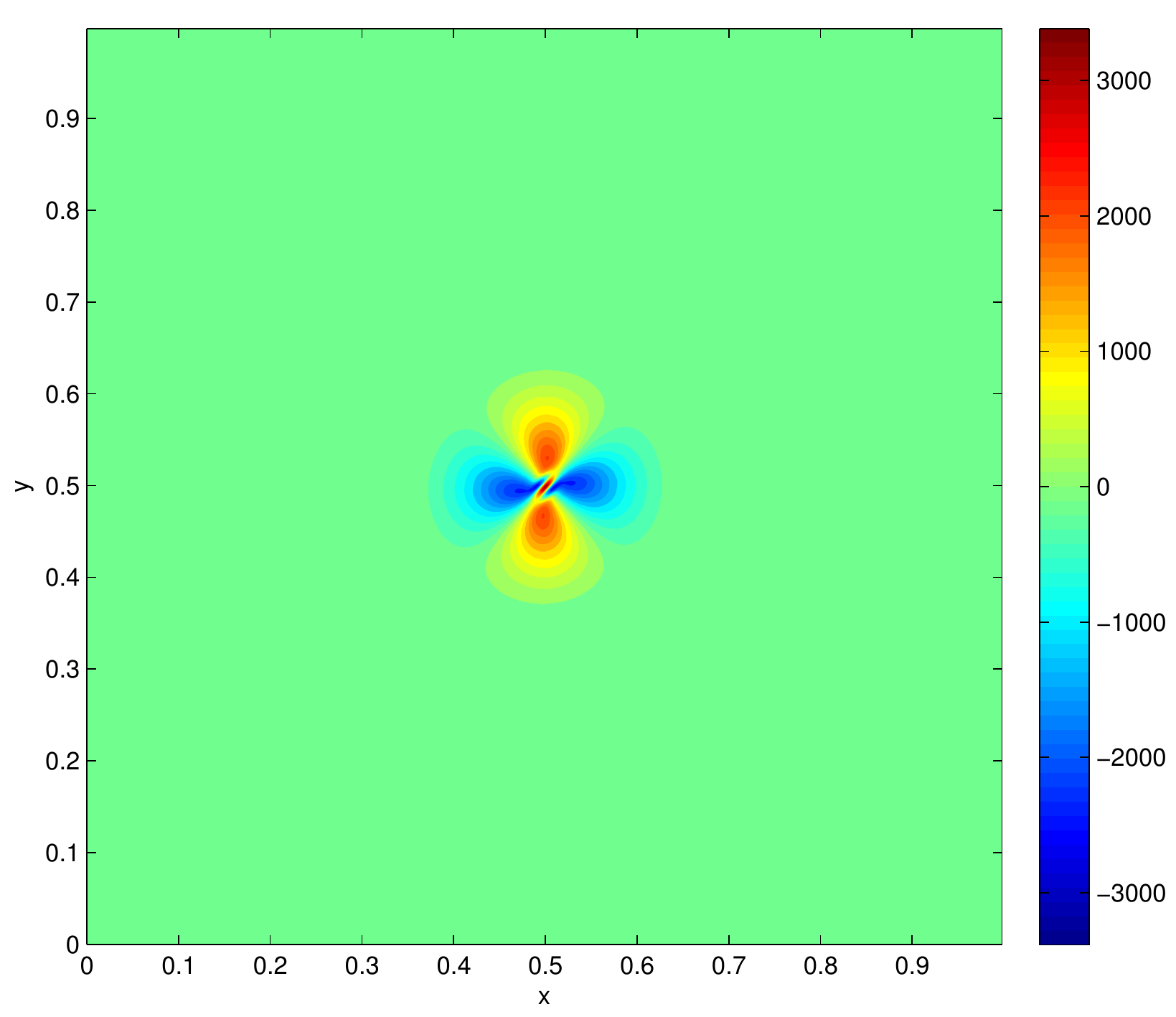}}\\
\subfigure[]{\includegraphics[width=0.3\textwidth]{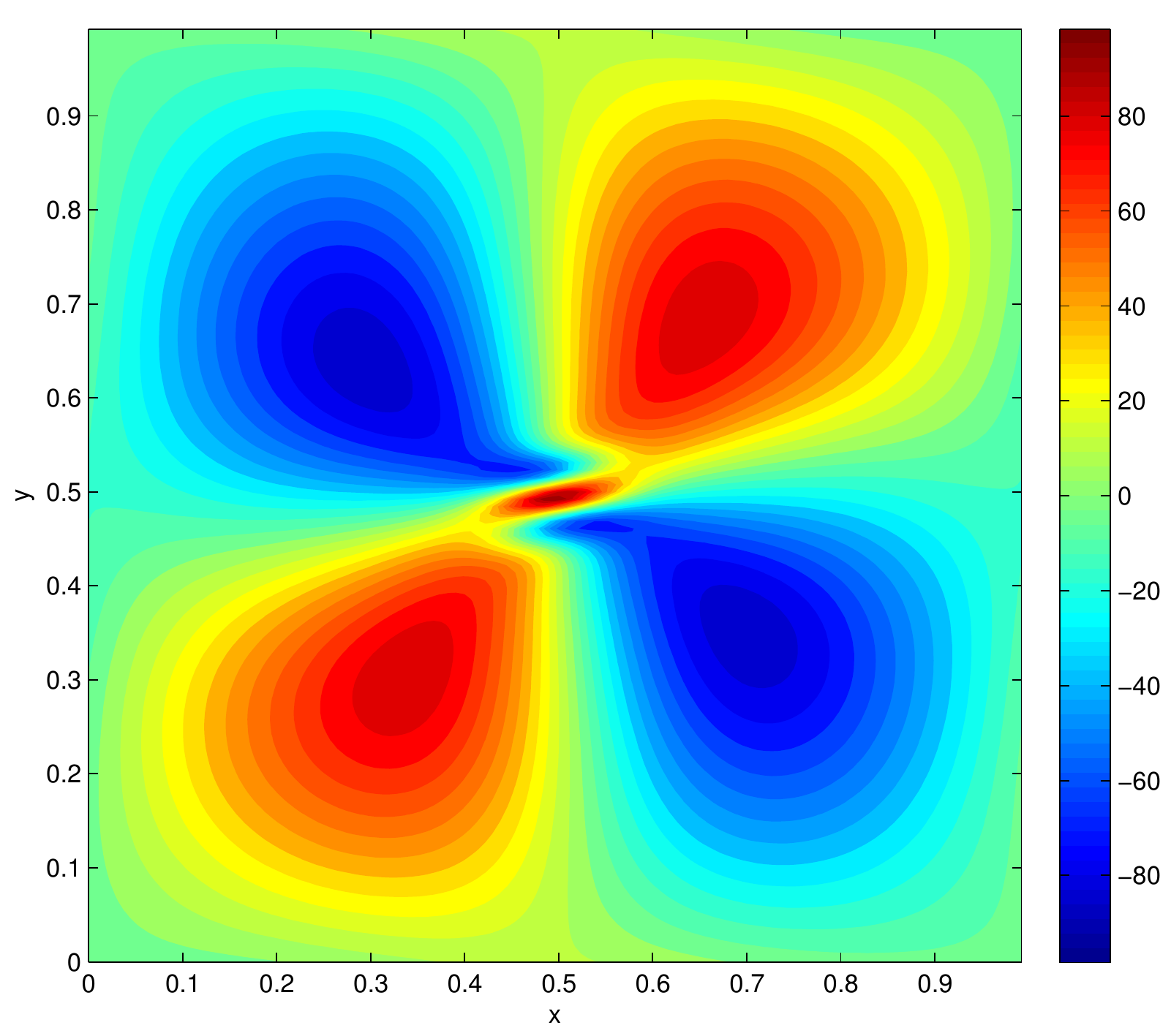}}
\subfigure[]{\includegraphics[width=0.3\textwidth]{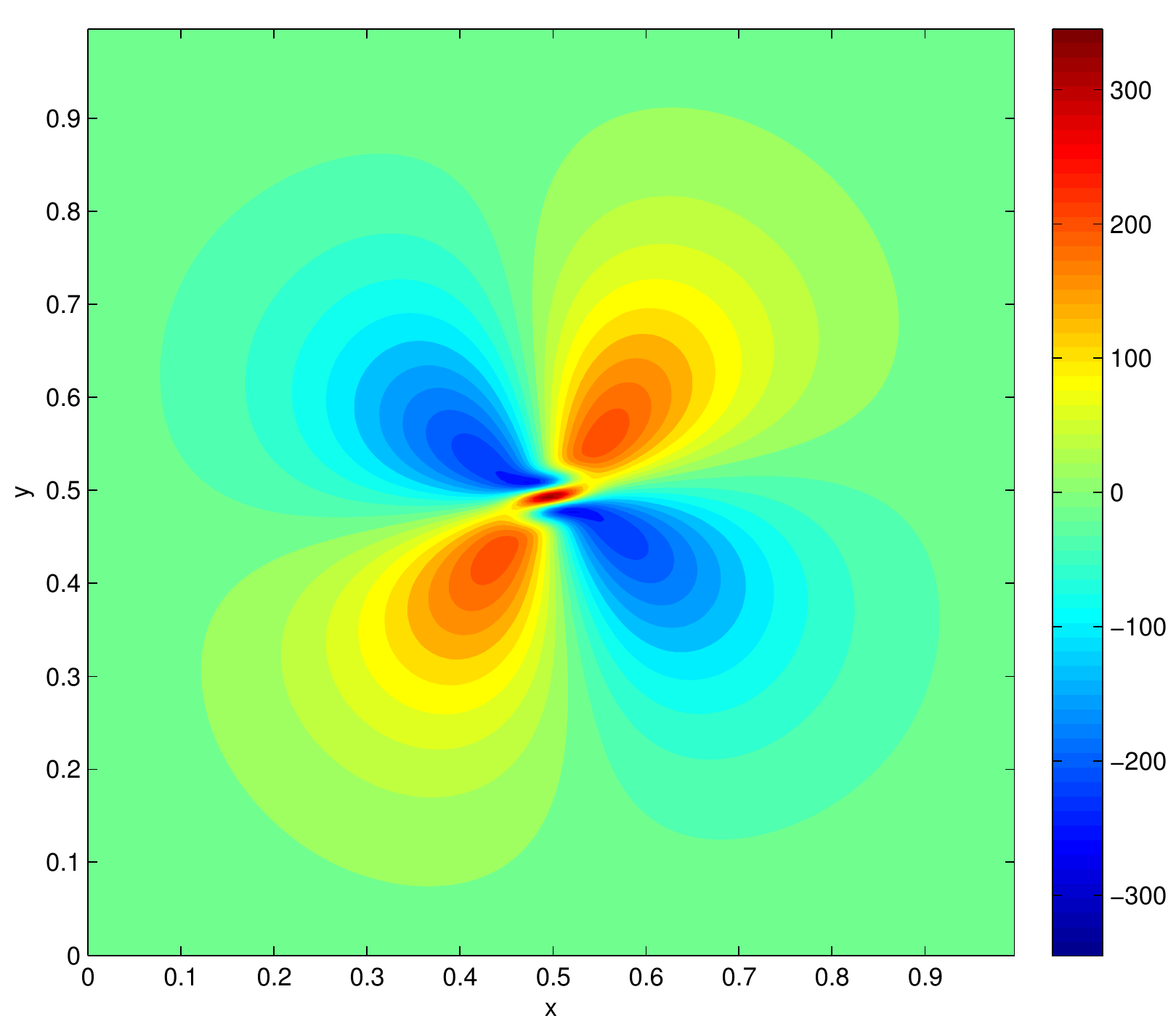}}
\subfigure[]{\includegraphics[width=0.3\textwidth]{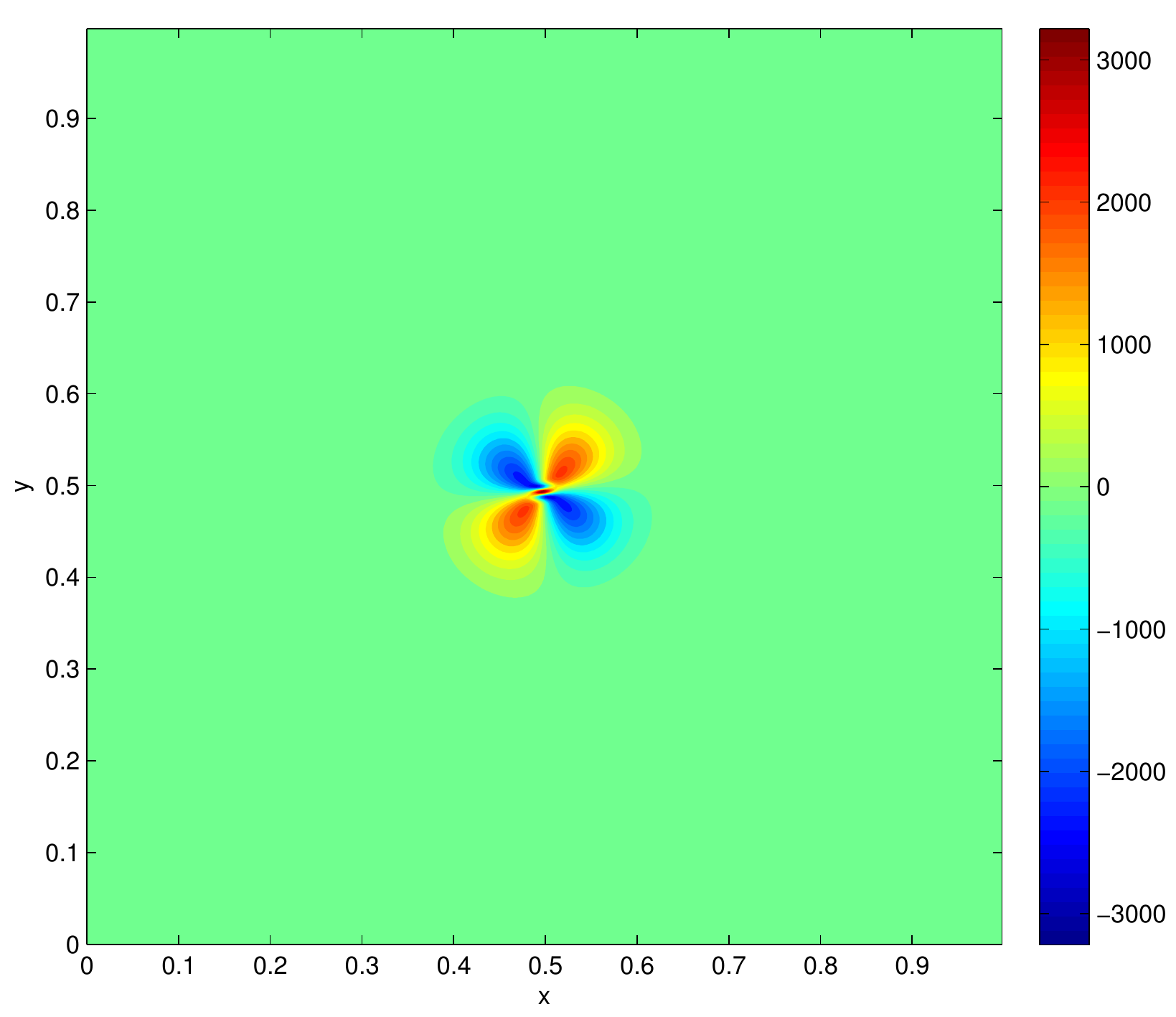}}
\caption{Dependence of the maximum palinstrophy rate of growth
  $\R_{\P_0}(\tpKP)$ on $\P_0$ for (a) $\K_0 = 10$ and (b) $\K_0 =
  10^0$ (solid), $\K_0 = 10^1$ (dashed) and $\K_0 = 10^2$
  (dash-dotted).  Figure (a) shows two solution branches, whereas
  figure (b) only the ones with larger values of $\R_{\P_0}$.  Optimal
  vortex states corresponding to the two branches, marked with the
  solid and dashed lines in figure (a), are shown in figures (c--e)
  and (f--h), respectively, for the following palinstrophy values:
  (c,f) $\P_0 = 10\P_c$, (d,g) $\P_0 = 10^2\P_c$ and (e,h) $\P_0 =
  10^4\P_c$ (marked with short vertical dashes), where $\P_c =
  (2\pi)^4\K_0$ is the Poincar\'{e} limit indicated with vertical
  dash-dotted lines in figures (a) and (b).}
\label{fig:K0P0}
\end{center}
\end{figure}

\begin{figure}
\begin{center}
  \includegraphics[width=0.4\textwidth]{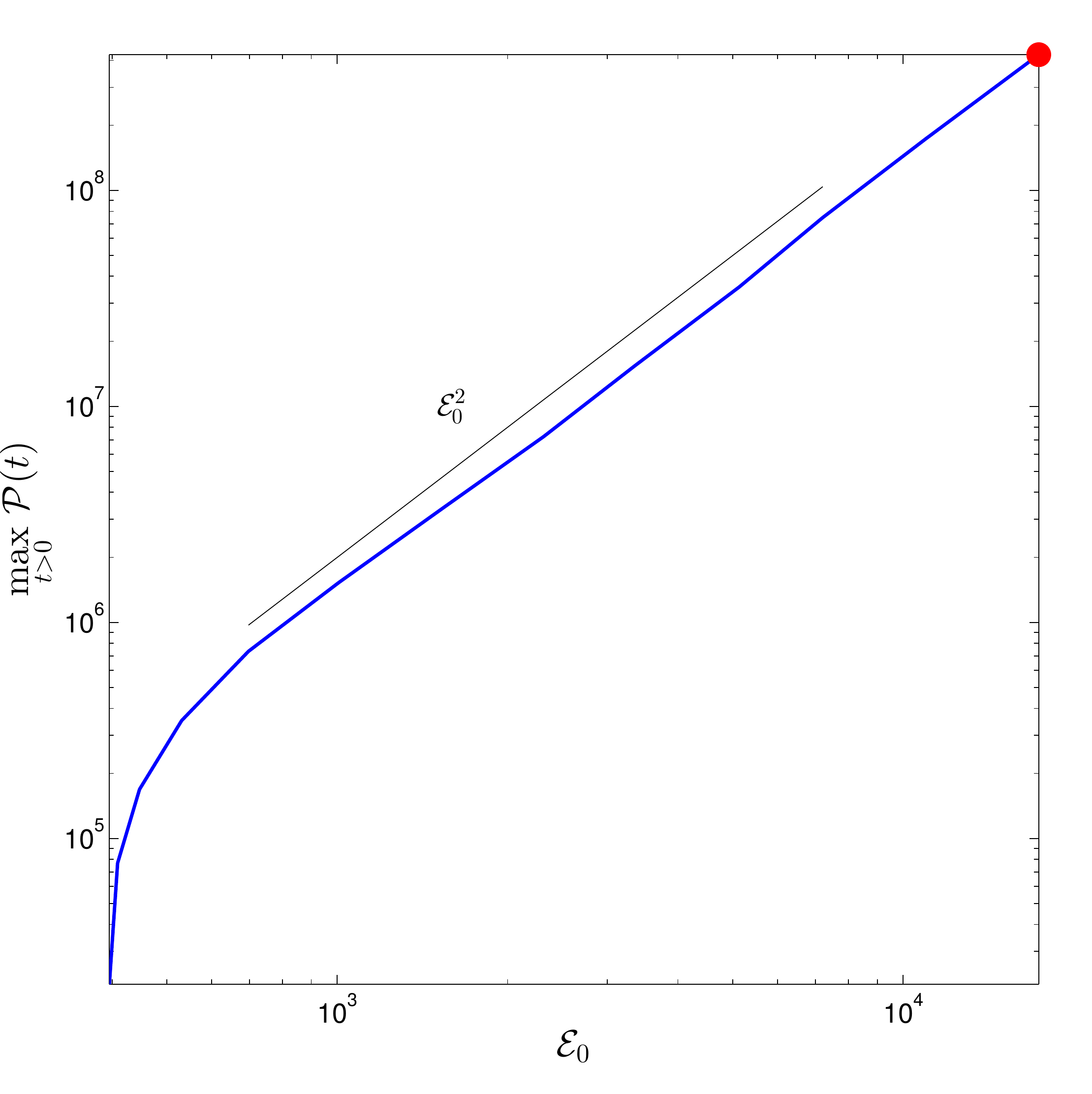}
  \caption{Maximum palinstrophy $\P_{\max}$ as a function of initial
    enstrophy $\E_0$ with the optimal vortex states from the upper
    branch in Figure \ref{fig:K0P0}(a) used as the initial data in the
    solution of Navier-Stokes system \eqref{eq:NS2Da}--\eqref{eq:NS2Dc}.
    Solid symbol (in the top right-hand corner) represents the flow
    whose time evolution is analyzed in detail in Sections
    \ref{sec:short} and \ref{sec:long}.}
\label{fig:maxPt_vsE0}
\end{center}
\end{figure}

For a fixed energy $\K_0$ and in the limit of small palinstrophies
$\P_0 \rightarrow (2\pi)^4 \, \K_0$, solutions of optimization problem
\eqref{eq:optR_KP} can in fact be found analytically and turn out to be
the eigenfunctions of the Laplacian operator. For intermediate and
large values of the palinstrophy the maximizers $\tpKP$ are found
numerically with a discretized gradient flow combining a variational
technique for the determination of the gradient of $\R_{\P}(\psi)$
with respect to $\psi$ and a special method to simultaneously enforce
the two constraints. As shown in Figure \ref{fig:K0P0}, for each value
of $\K_0$ two families of maximizing vortex states parameterized by
$\P_0$ were found corresponding to the staggered and aligned
arrangement of the vortex cells. We note that in the limit of large
$\P_0$ the rate of growth of palinstrophy characterizing these
families exhibits a clear power-law
\begin{equation}
\frac{d\P}{dt} \sim \P_0^{1.49 \pm 0.02}
\label{eq:dPdt_KP_num}
\end{equation}
demonstrating that estimate \eqref{eq:dPdt_KP} is in fact sharp (up to
a numerical prefactor). In other words, the families of vortex states
shown in Figures \ref{fig:K0P0}(c--h) exhibit the highest rate of
palinstrophy production allowed by the mathematical analysis of 2D
Navier-Stokes system. An intriguing feature of this family of
maximizing vortex states is that, as shown in Figure
\ref{fig:maxPt_vsE0}, the time evolution starting from these fields as
the initial data \eqref{eq:NS2Dc} also saturates finite-time estimate
\eqref{eq:maxPt_Ayala}. Although the maximizing states $\tpKP$ are
obtained with fixed energy $\K_0$ and palinstrophy $\P_0$,
cf.~\eqref{eq:optR_KP}, to be consistent with the right-hand side of
estimate \eqref{eq:maxPt_Ayala}, the data in Figure
\ref{fig:maxPt_vsE0} is plotted with the corresponding enstrophy
$\E_0$ on the abscissa. We emphasize that, in contrast to $\tpKP$, the
maximal vortex states obtained under a single constraint on $\P_0$, or
by fixing $\E_0$ and $\P_0$, {\em did not} saturate the finite time
estimates \cite{ap13a}.  Thus, the number and choice of the constrains
imposed when solving this type of optimization problems is quite
important and deserves further study in the context of the 1D Burgers
and 3D Navier-Stokes systems. In the next Section we analyze in detail
an example of such an extreme time-evolution of the vorticity field
from the family saturating estimate \eqref{eq:maxPt_Ayala}.

\section{Finite-Time Evolution of Instantaneously Optimal Initial Data}
\label{sec:ft}

Our main goal in this Section is to identify the physical mechanisms
responsible for the growth of the palinstrophy over finite-time
intervals which saturates estimate \eqref{eq:maxPt_Ayala}. To this end,
we select one representative case (marked with a solid symbol) from
the family shown in Figure \ref{fig:maxPt_vsE0} and follow its
evolution in time, first over a short-time window (in subsection
\ref{sec:short}) and then over a longer time-window (in subsection
\ref{sec:long}). Navier-Stokes system \eqref{eq:NS2Da}--\eqref{eq:NS2Dc}
is solved using $-\Delta\tpKP$ with $\K_0 = 10$ and $\P_0 = 1.714\cdot
10^8$ (which correspond to $\E_0 = 1.735\cdot 10^4$ in Figure
\ref{fig:maxPt_vsE0}) as the initial data. While the maximum
palinstrophy value $\P_{\max}$ is achieved at short times, the
long-time evolution is also shown here as it reveals a number of
interesting aspects. The short-time evolution of all cases shown in
Figure \ref{fig:maxPt_vsE0} with $\E_0$ (or, equivalently, $\P_0$)
sufficiently large follows essentially the same scenario as described
below. On the other hand, some details of the long-time evolution may
differ between the different cases. Navier-Stokes system
\eqref{eq:NS2Da}--\eqref{eq:NS2Dc} is solved numerically with an
approach combining a Krylov subspace method for the time
discretization with a standard pseudo-spectral technique for the
discretization in space. The spatial resolution of $2048^2$ ensures
that the evolution studied below is well resolved.

\subsection{Short-Time Evolution}
\label{sec:short}

\begin{figure}
\centering
\mbox{\subfigure[]{
\Bmp{0.5\textwidth}\vspace*{-0.75cm}
\includegraphics[width=1.0\textwidth]{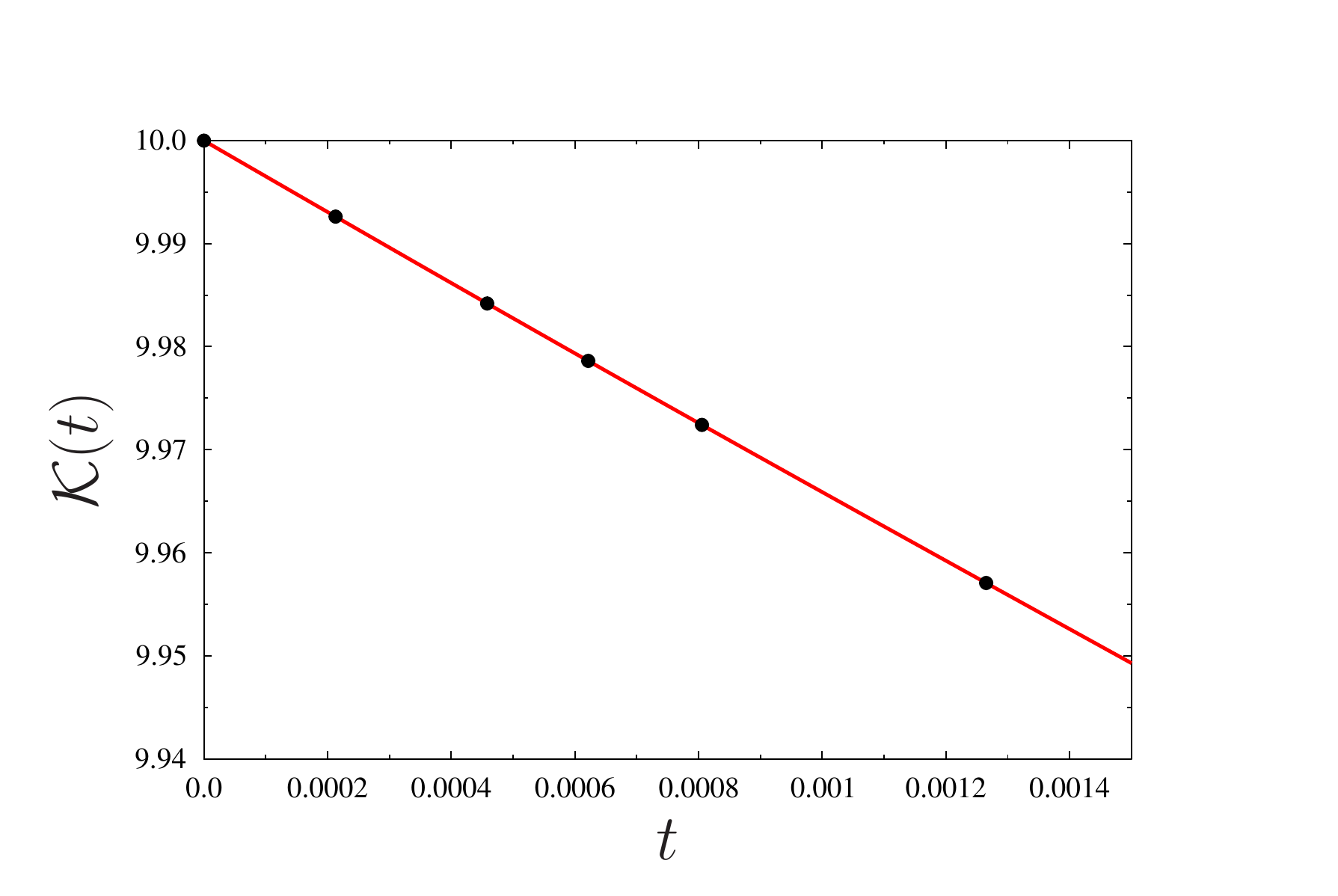}\Emp} 
\subfigure[]{
\Bmp{0.5\textwidth}\vspace*{-0.75cm}
\includegraphics[width=1.0\textwidth]{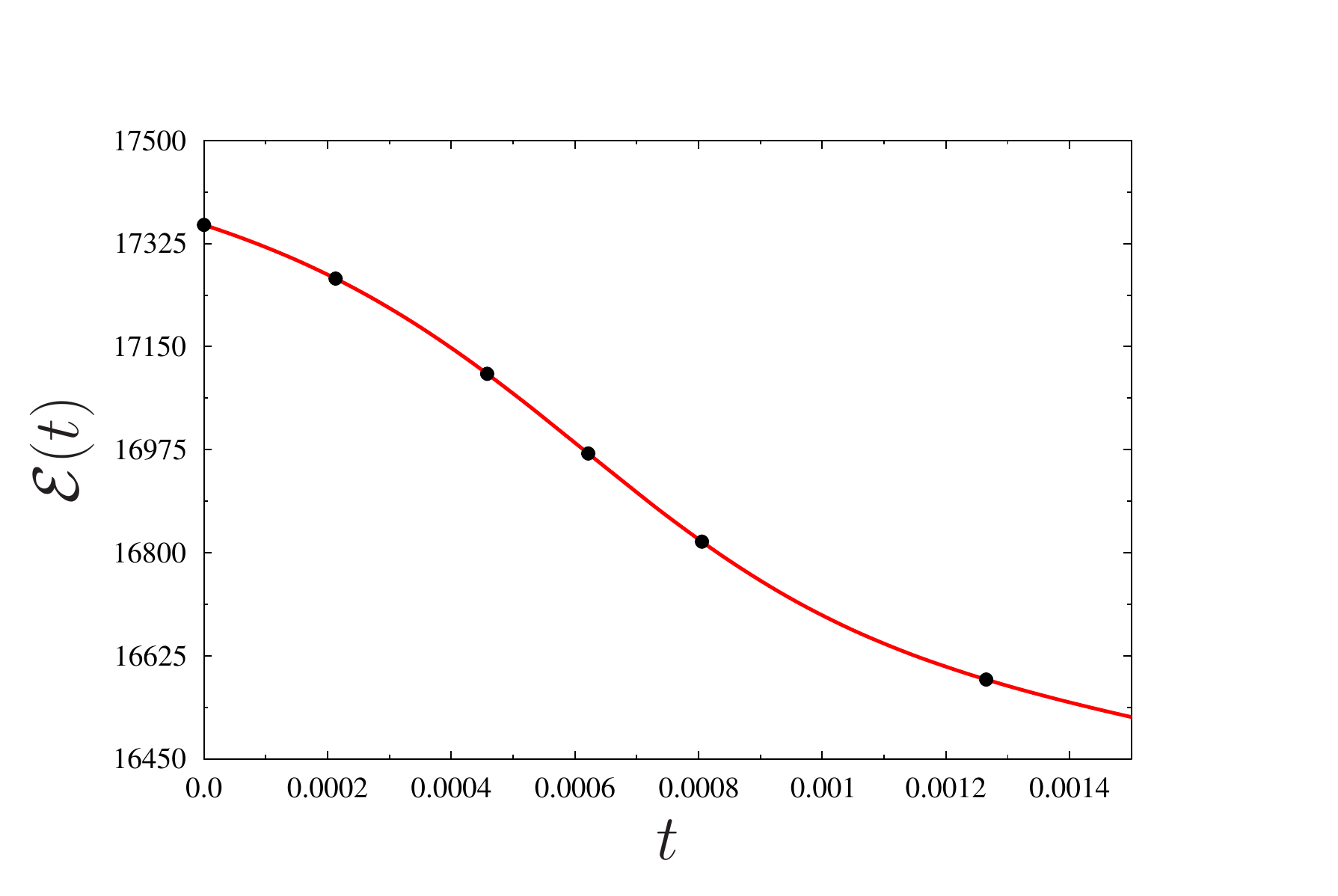}\Emp}}
\subfigure[]{
\Bmp{0.5\textwidth}\vspace*{-0.75cm}
\includegraphics[width=1.0\textwidth]{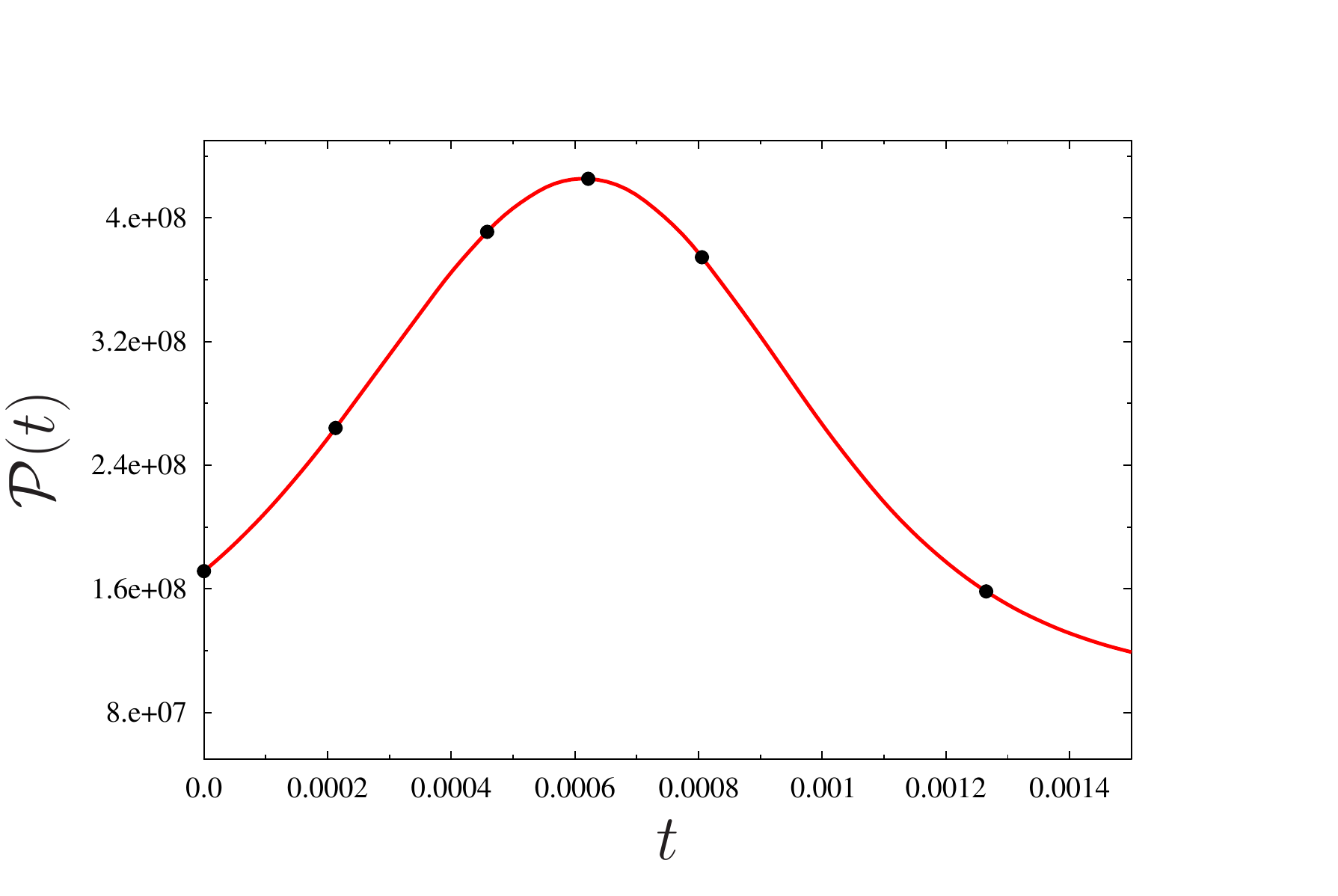}\Emp} 
\caption{Time histories of (a) energy $\K(t)$, (b) enstrophy $\E(t)$
  and (c) palinstrophy $\P(t)$ during an initial stretching event.
  Solid symbols represent the instances of time for which the
  vorticity fields are shown in Figure \ref{fig:vor1}.}
\label{fig:kep1}
\end{figure}

\begin{figure}
\centering
\mbox{
\subfigure[$t=0.0$]{
\includegraphics[width=0.4\textwidth]{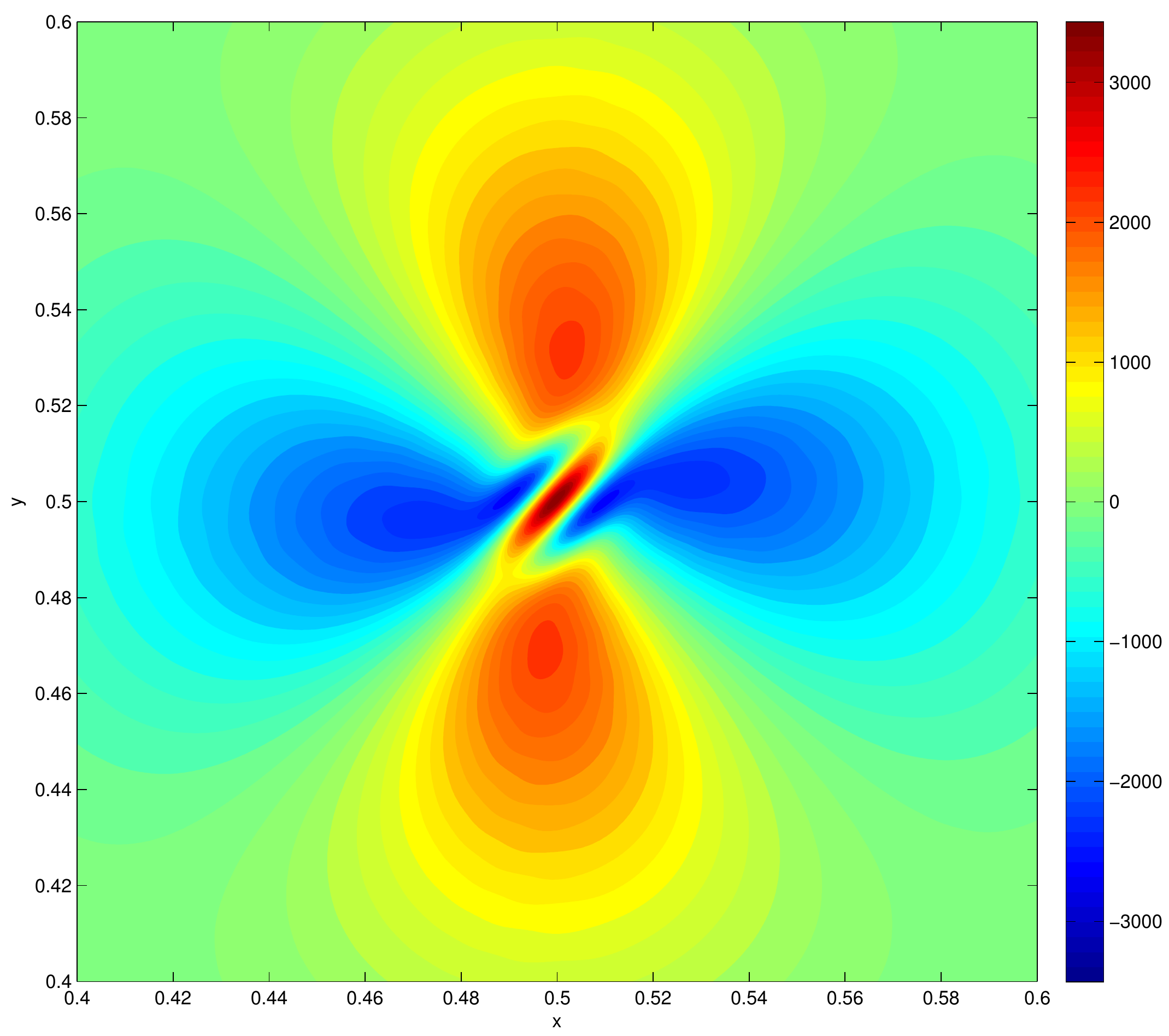}}\qquad
\subfigure[$t=0.000213$]{
\includegraphics[width=0.4\textwidth]{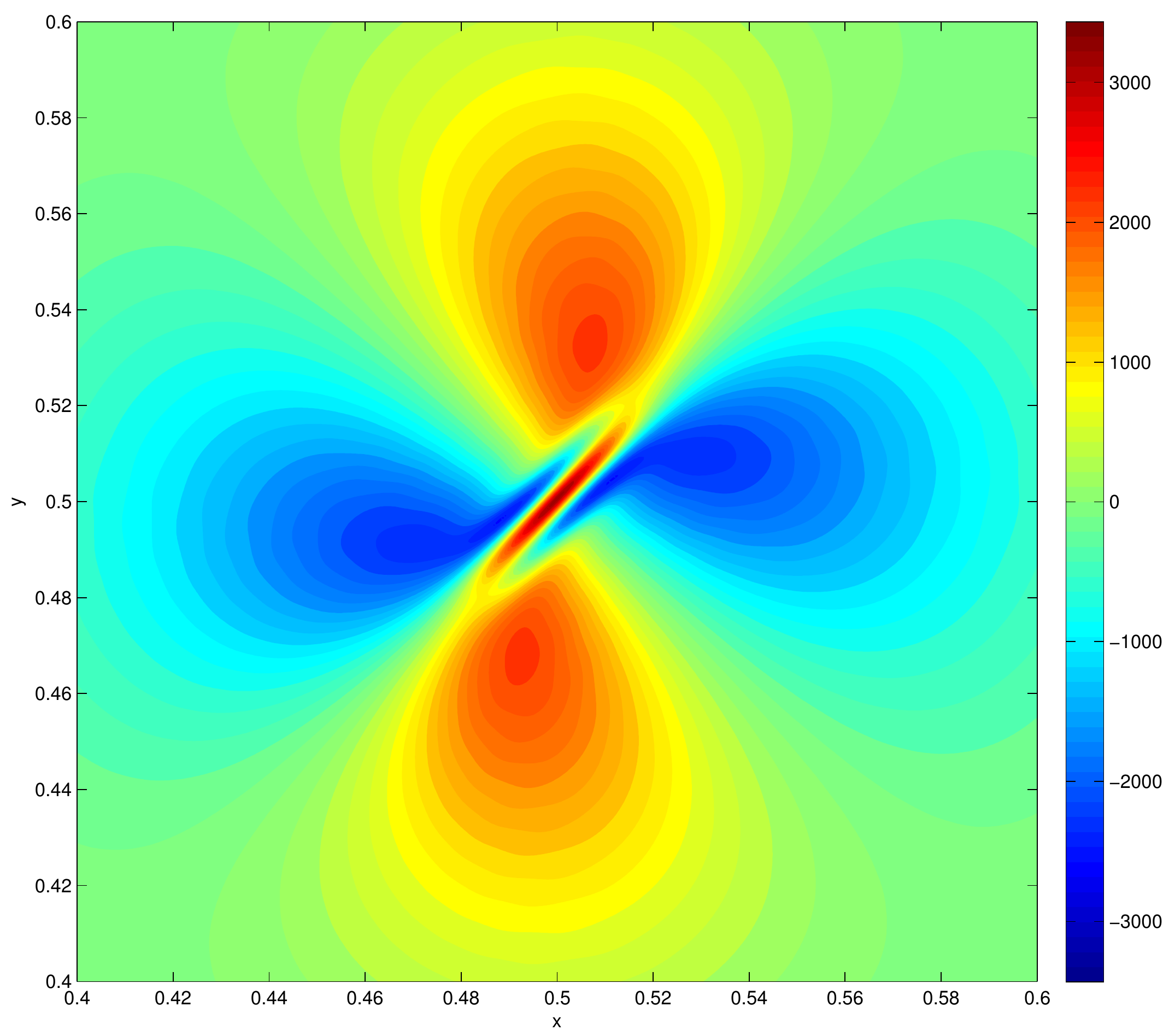}}}
\mbox{
\subfigure[$t=0.000458$]{
\includegraphics[width=0.4\textwidth]{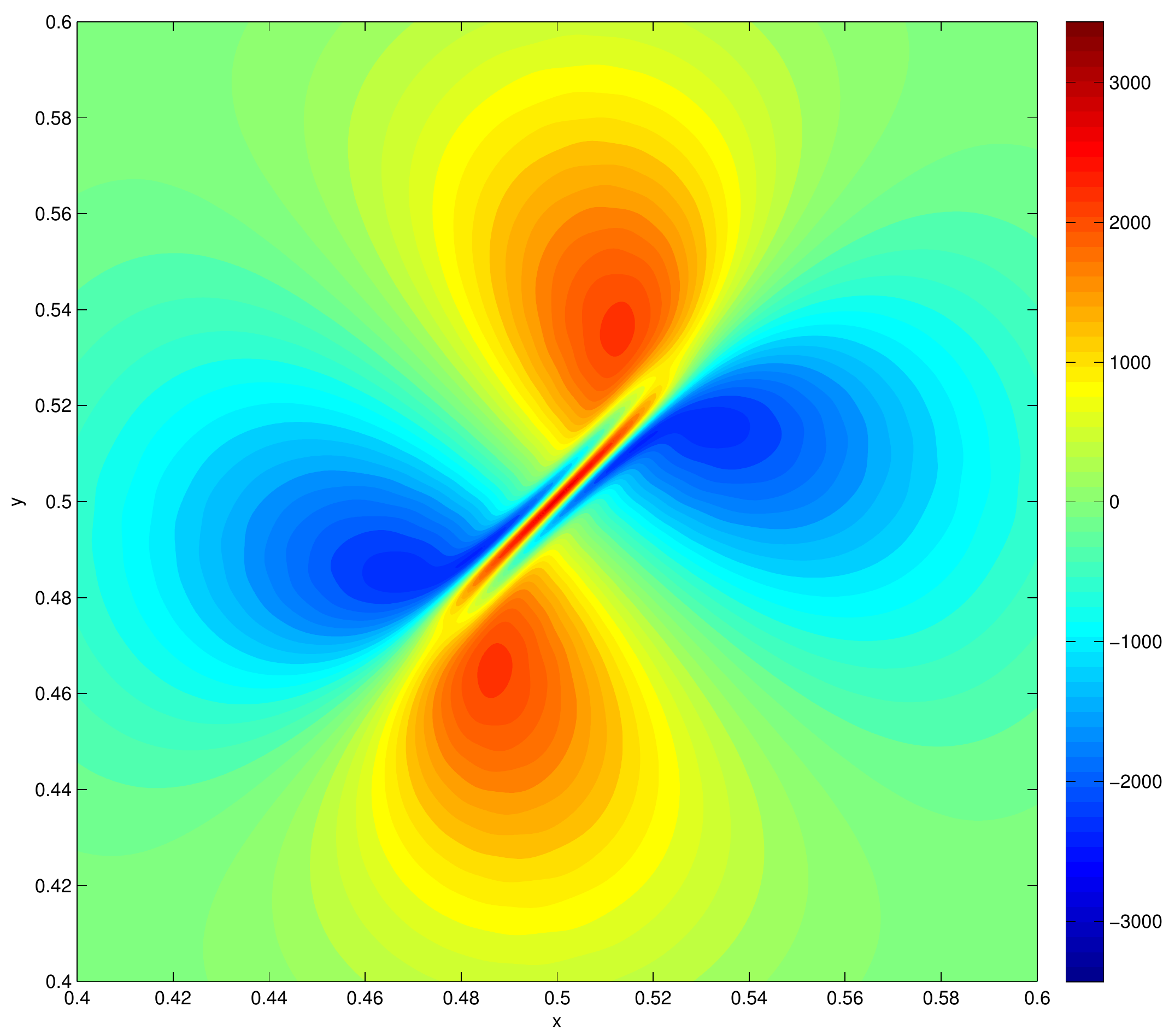}}\qquad
\subfigure[$t=0.000633$]{
\includegraphics[width=0.4\textwidth]{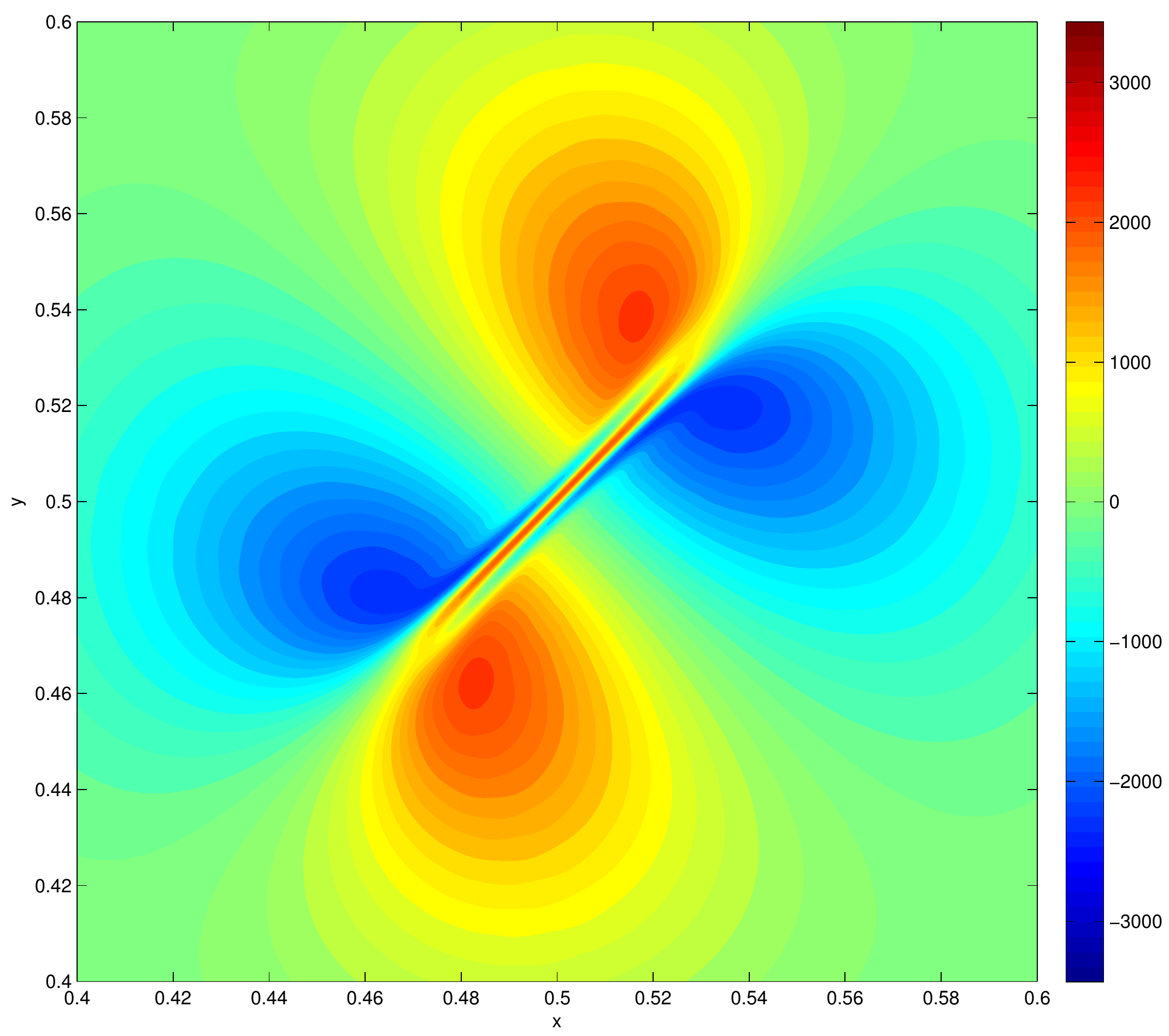}}}
\mbox{
\subfigure[$t=0.000805$]{
\includegraphics[width=0.4\textwidth]{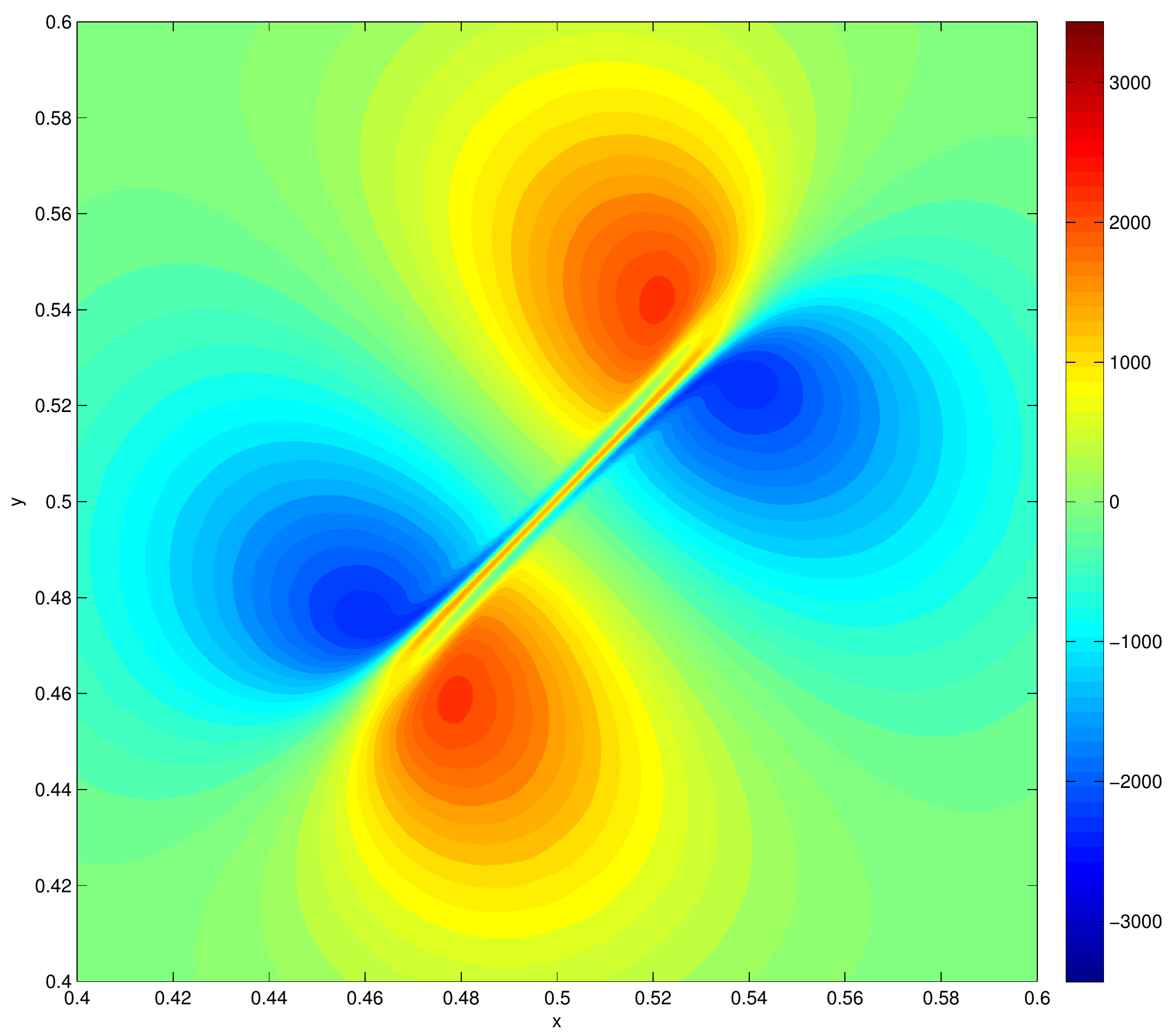}}\qquad
\subfigure[$t=0.001265$]{
\includegraphics[width=0.4\textwidth]{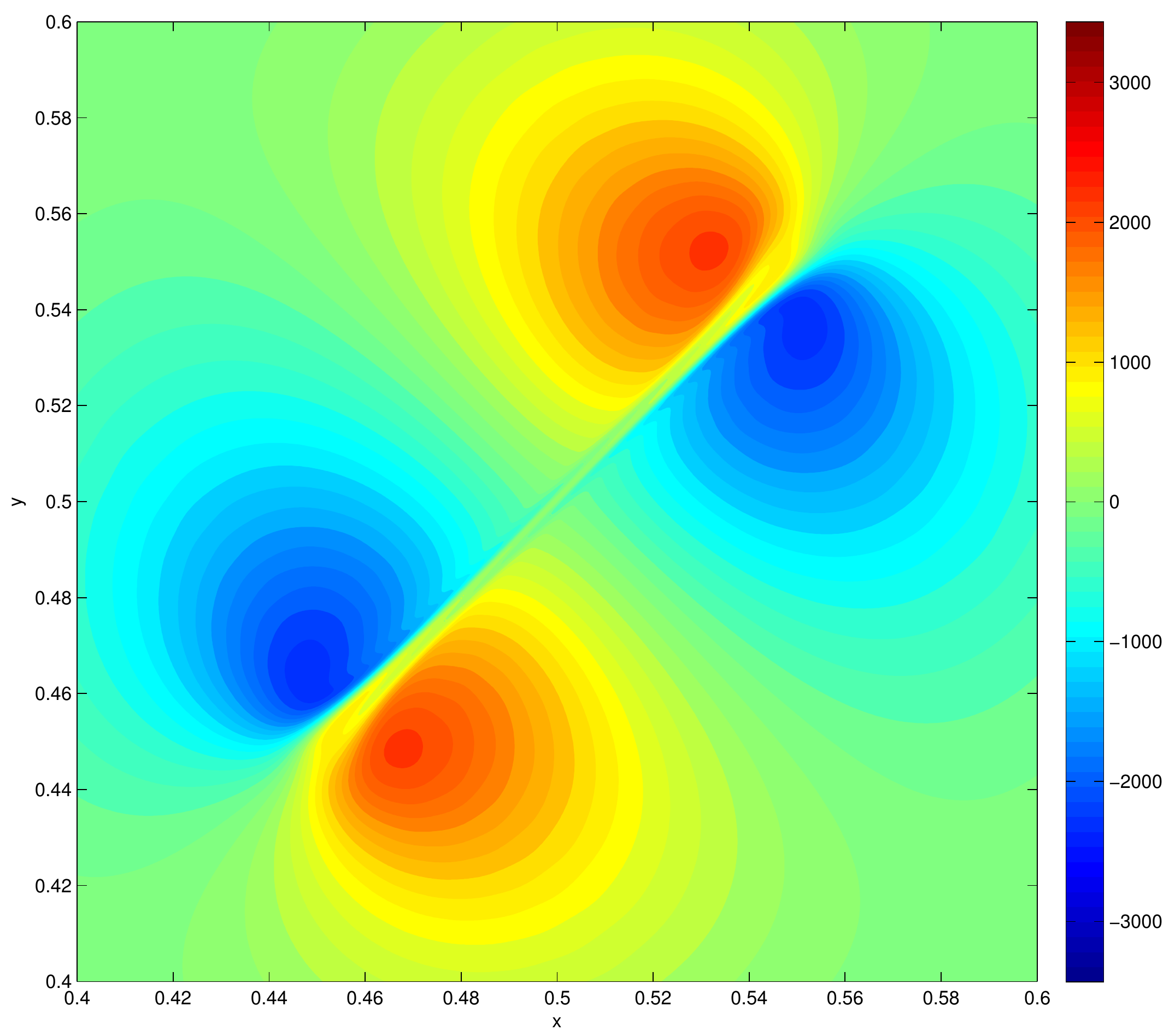}}}
\caption{Vorticity fields at the times indicated, and marked with
  solid symbols in Figures \ref{fig:kep1}(a,b,c), during an initial
  stretching event. Field (a) is the instantaneously optimal vorticity
  distribution $-\Delta\tpKP$, cf.~\eqref{eq:optR_KP}, with $\K_0 = 10$
  and $\P_0 = 1.714 \cdot 10^{8}$ used as the initial data
  \eqref{eq:NS2Dc} in the Navier-Stokes system.}
\label{fig:vor1}
\end{figure}

We analyze here the ``stretching event'' occurring at the beginning of
the time evolution. The history of energy $\K(t)$, enstrophy $\E(t)$
and palinstrophy $\P(t)$ is shown in Figures \ref{fig:kep1}(a,b,c).
The snapshots of the vorticity field $\omega(t)$ at some
representative instances of time during the event (marked with solid
symbols in the plots in Figures \ref{fig:kep1}(a,b,c)) are shown in
Figure \ref{fig:vor1}. As discussed in Section \ref{sec:maxdpdt}, the
palinstrophy increase $(\P_{\max} - \P_0)$ in terms of the initial
enstrophy $\E_0$ is as large as allowed by estimate
\eqref{eq:maxPt_Ayala}. This increase is clearly visible in Figure
\ref{fig:kep1}(c) and, in view of the relation $d\E(t) / dt = - 2 \nu
\P(t)$, is accompanied by accelerated dissipation of the enstrophy
visible in Figure \ref{fig:kep1}(b). The evolution of the vorticity
field, starting with the instantaneously optimal state $-\Delta\tpKP$
shown in Figure \ref{fig:vor1}(a), reveals the development of a thin
vortex filament stretched by the four satellite vortices. The state
when the peak palinstrophy value $\P_{\max}$ is reached is captured in
Figure \ref{fig:vor1}(d). After that, the central filament is
dissipated and the satellite vortices start to move apart under their
own induction as two slightly asymmetric dipoles, see Figure
\ref{fig:vor1}(e). An animation showing the short-time evolution of
the vorticity field is available on-line as supplementary material.

\subsection{Long-Time Evolution}
\label{sec:long}

\begin{figure}
\centering
\mbox{
\subfigure[]{
\Bmp{0.5\textwidth}\vspace*{-0.75cm}
\includegraphics[width=1.0\textwidth]{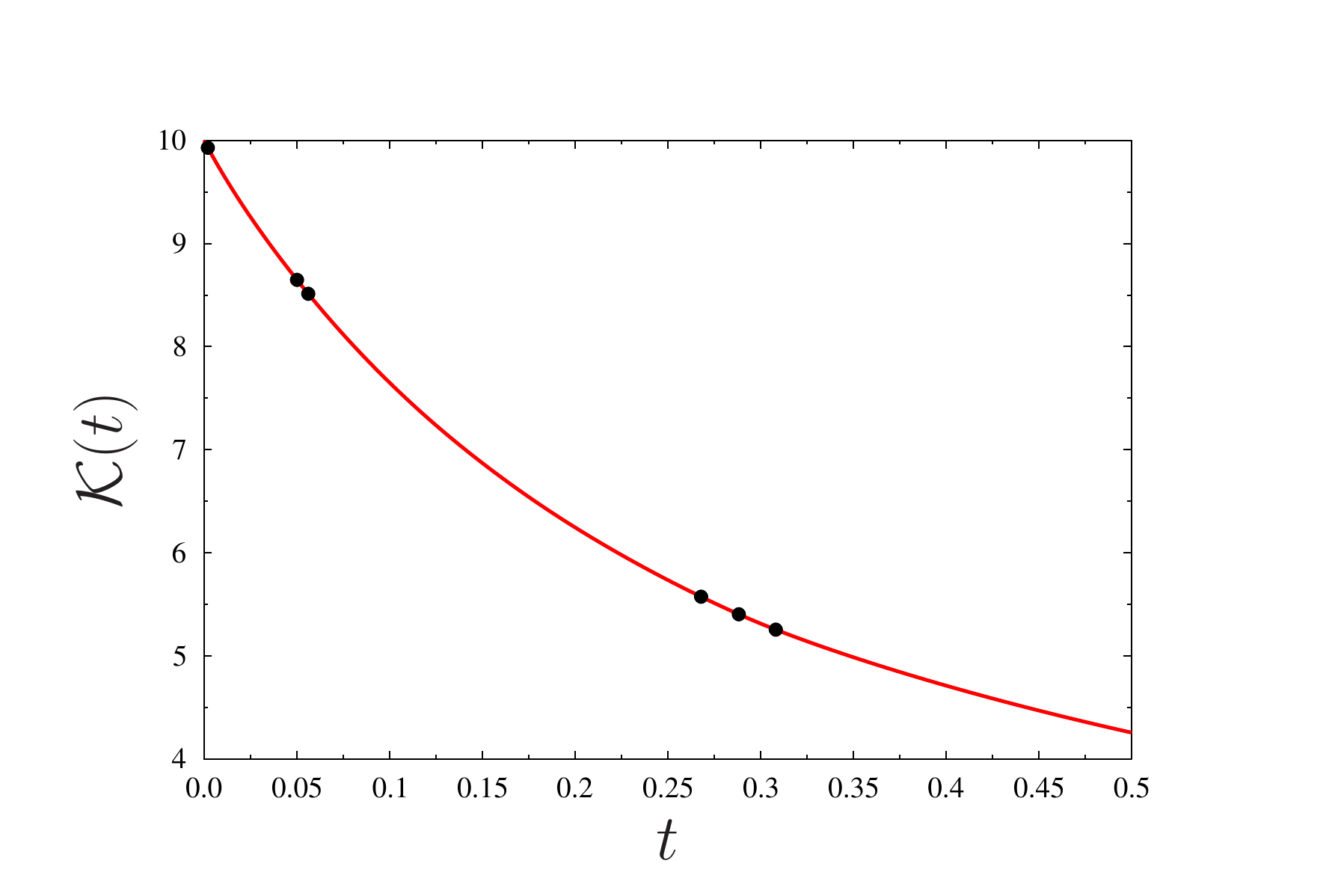}\Emp} 
\subfigure[]{
\Bmp{0.5\textwidth}\vspace*{-0.75cm}
\includegraphics[width=1.0\textwidth]{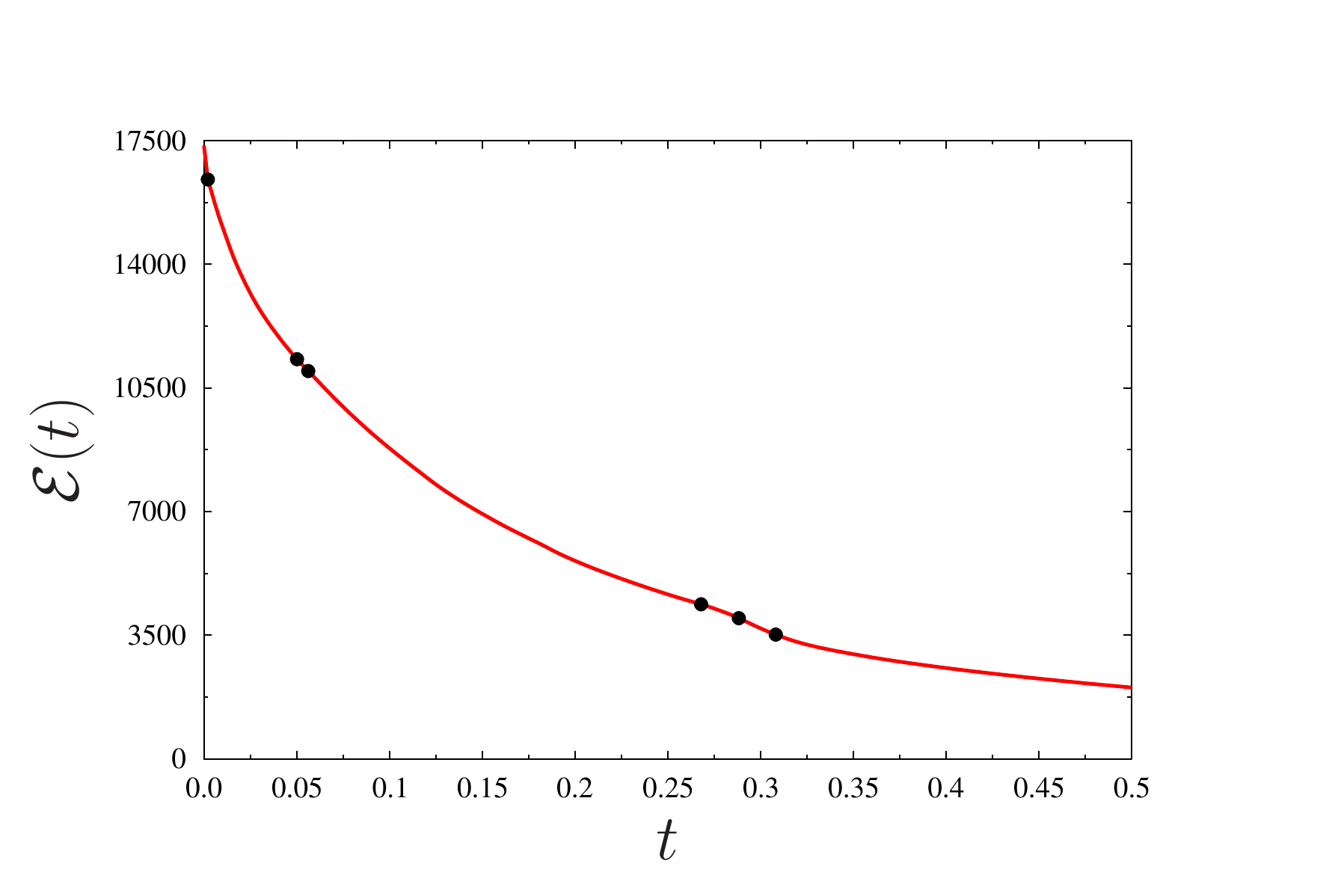}\Emp}}
\subfigure[]{
\Bmp{0.5\textwidth}\vspace*{-0.75cm}
\includegraphics[width=1.0\textwidth]{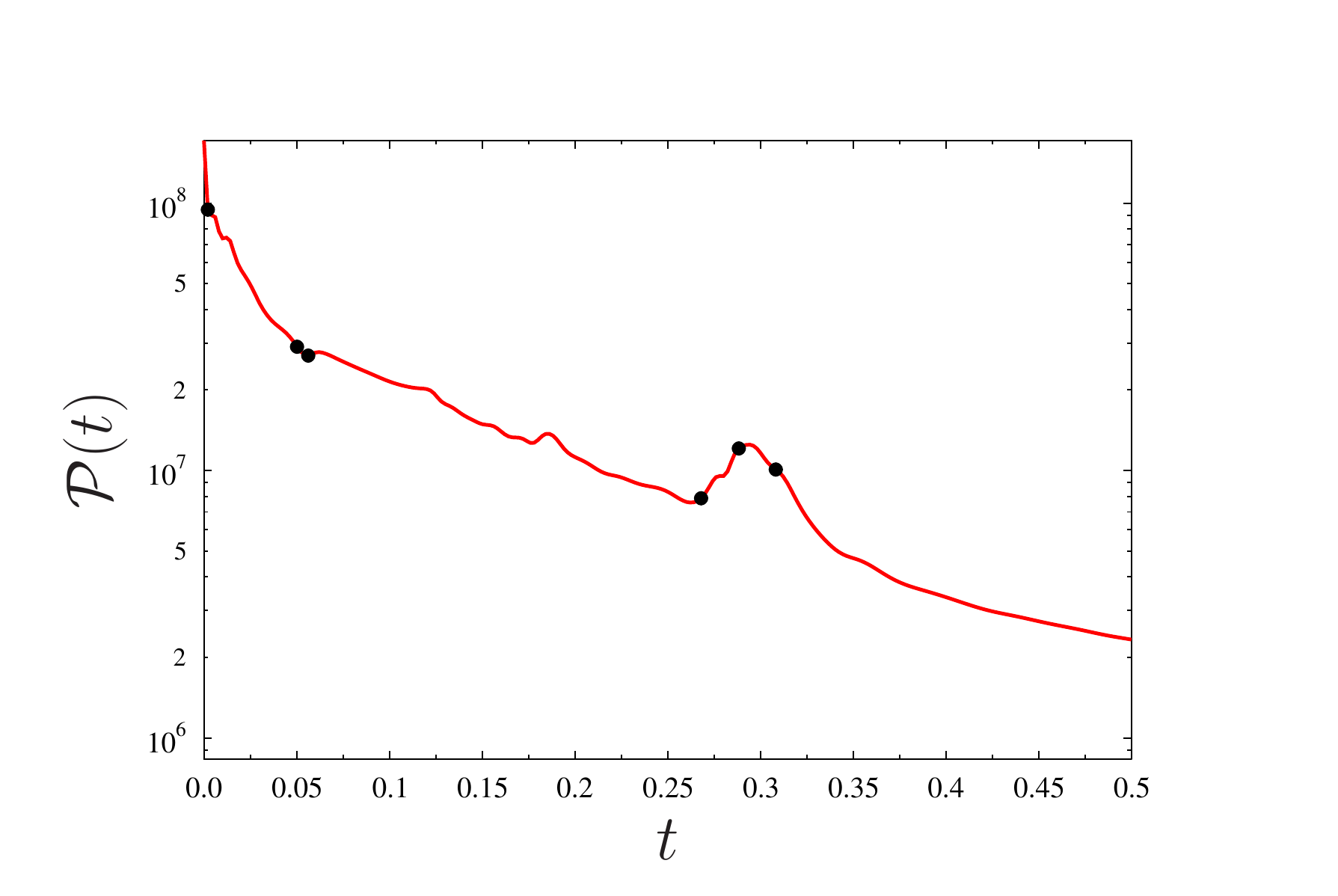}\Emp} 
\caption{Time histories of (a) energy $\K(t)$, (b) enstrophy $\E(t)$
  and (c) palinstrophy $\P(t)$ during the long-time evolution.
  Solid symbols represent the instances of time for which the
  vorticity fields are shown in Figure \ref{fig:vor2}.}
\label{fig:kep2}
\end{figure}

\begin{figure}
\centering
\mbox{
\subfigure[$t=0.00209$]{
\includegraphics[width=0.4\textwidth]{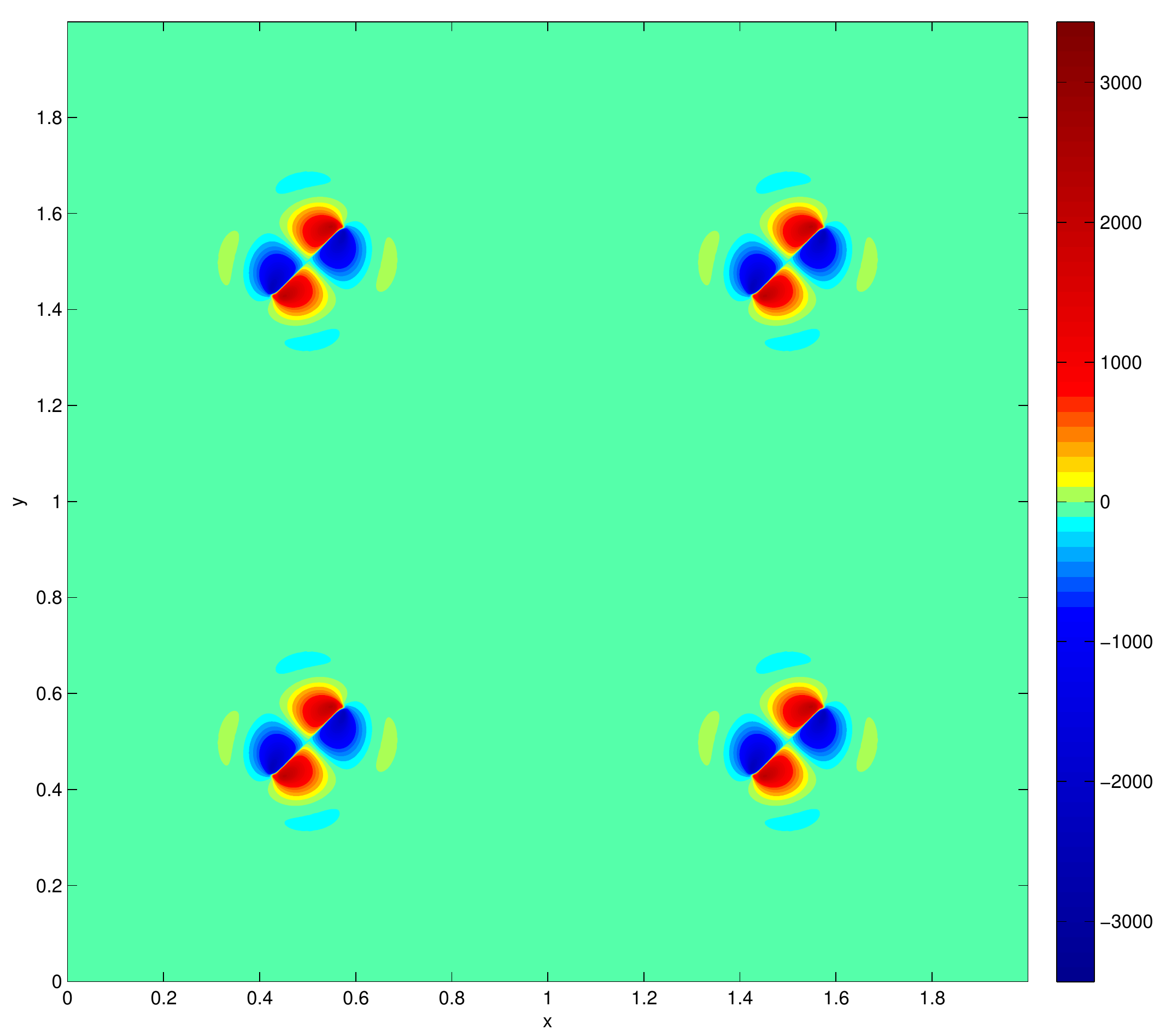}}\qquad
\subfigure[$t=0.05017$]{
\includegraphics[width=0.4\textwidth]{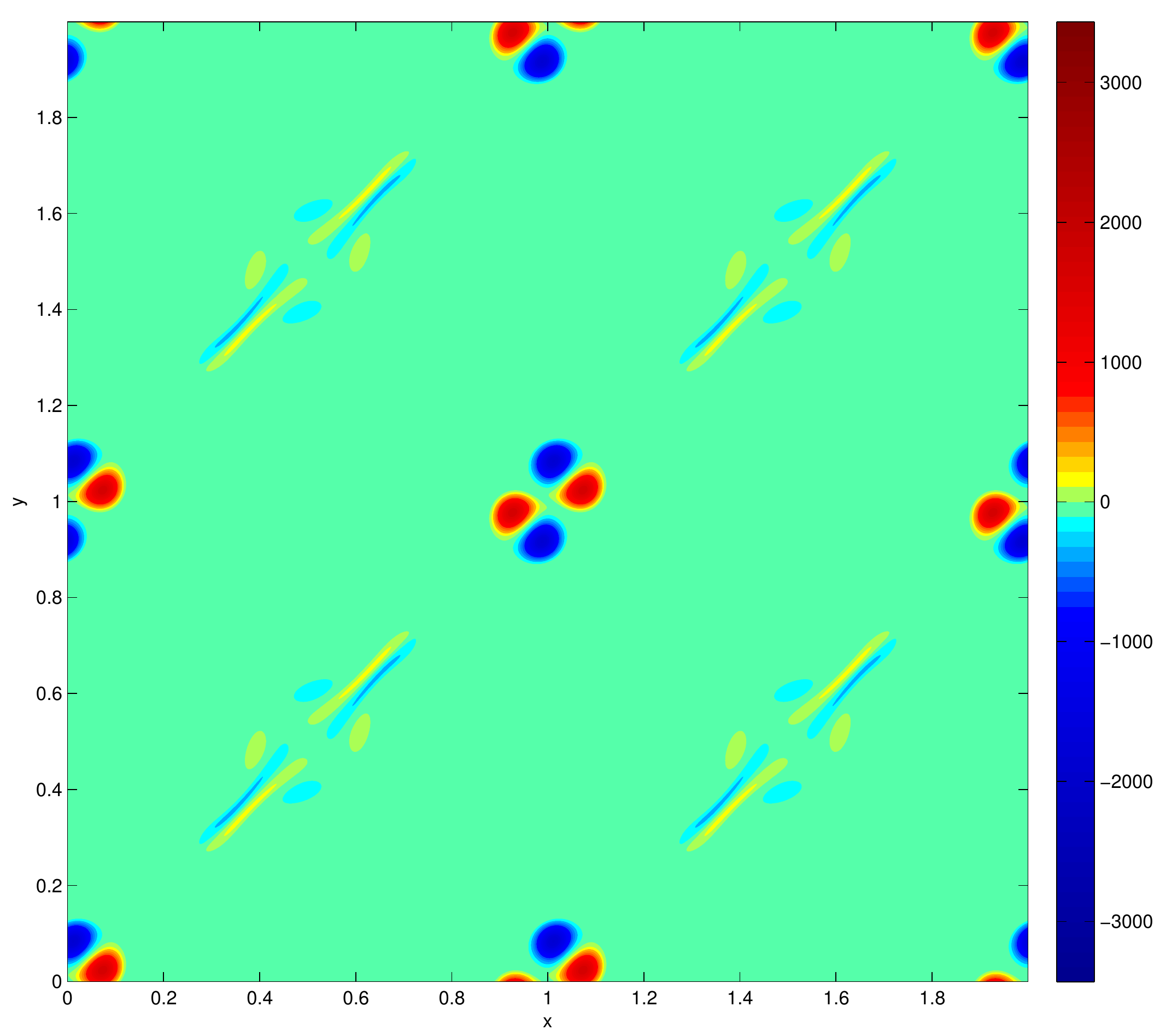}}}
\mbox{
\subfigure[$t=0.05623$]{
\includegraphics[width=0.4\textwidth]{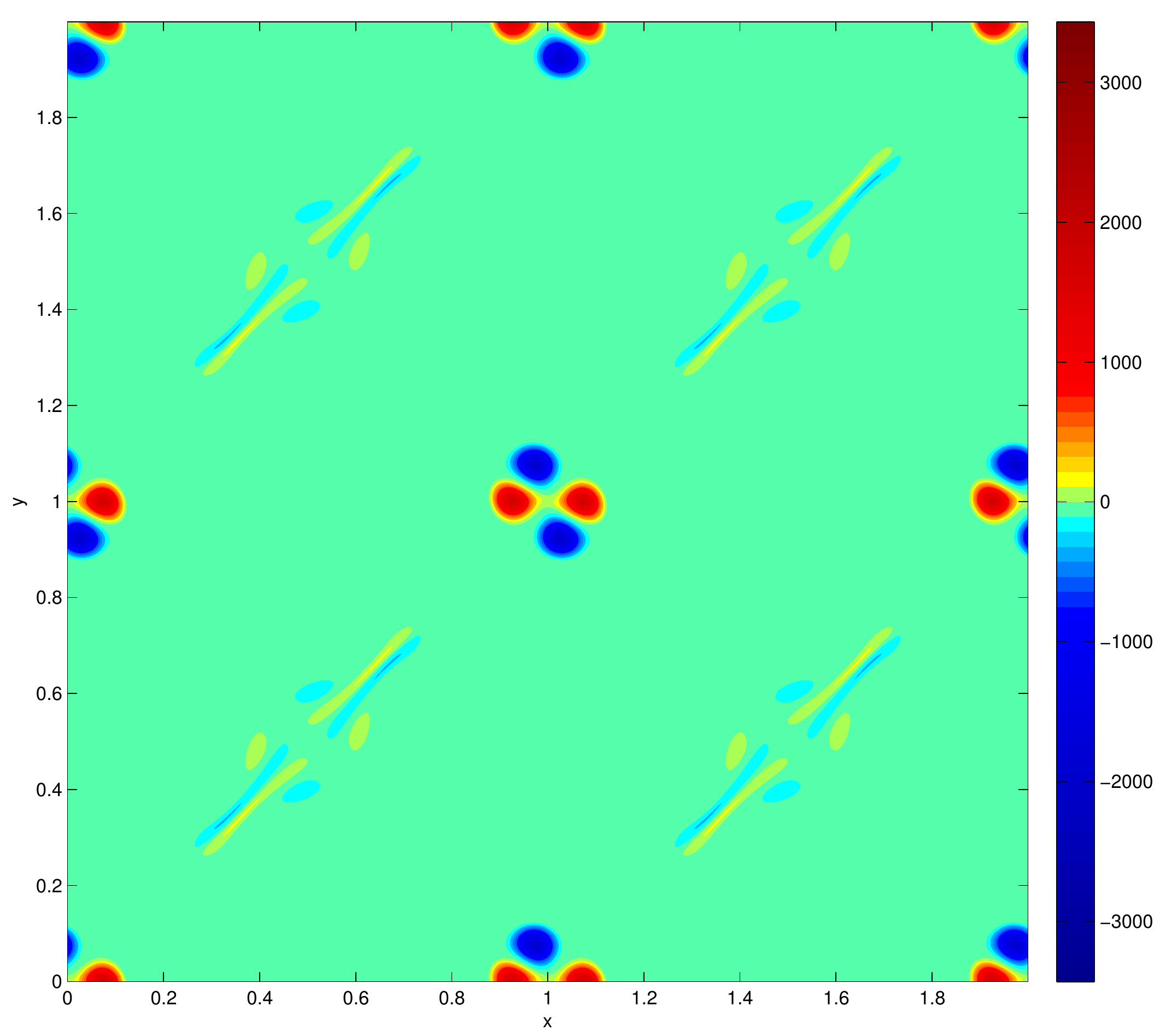}}\qquad
\subfigure[$t=0.26800$]{
\includegraphics[width=0.4\textwidth]{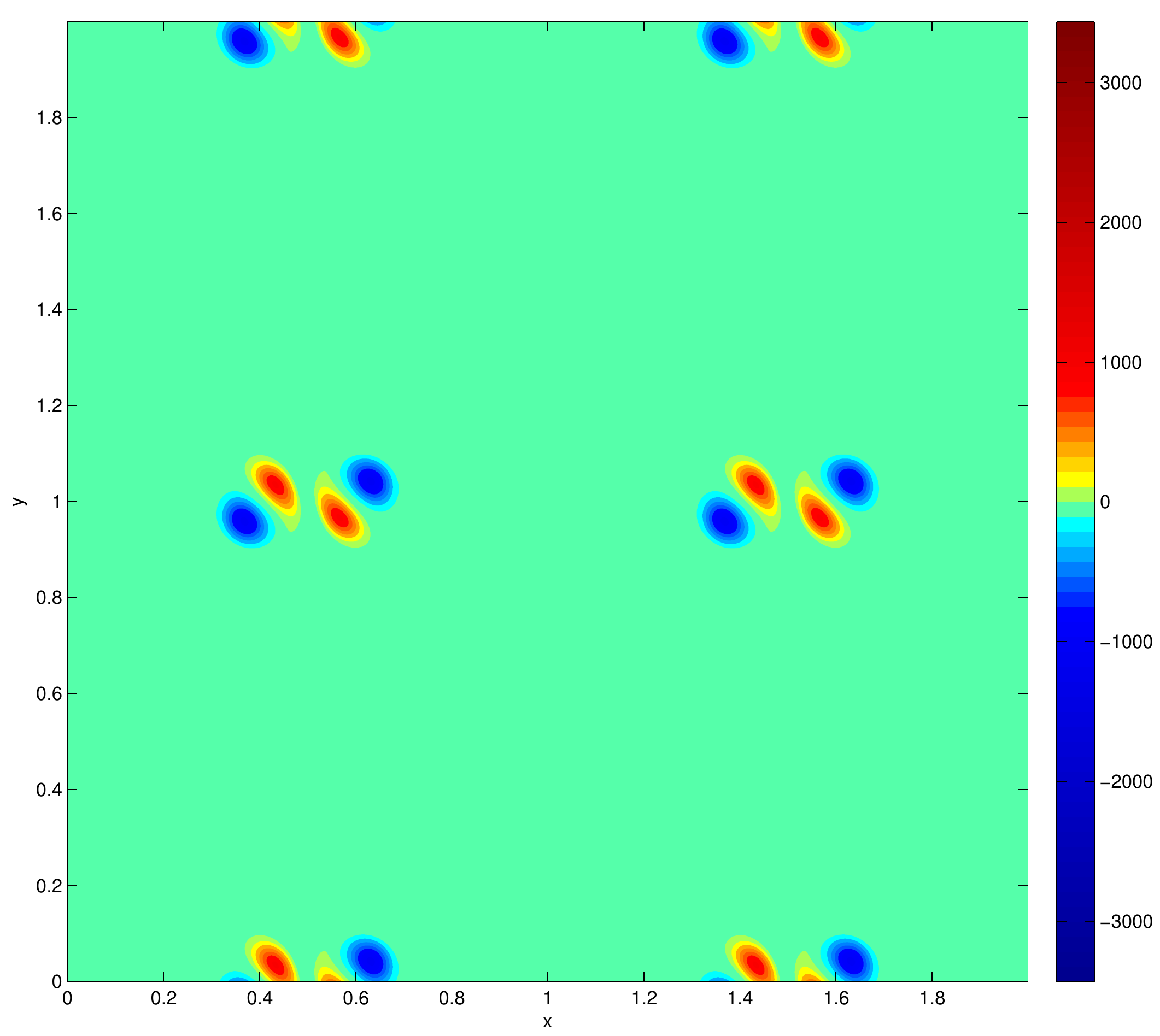}}}
\mbox{
\subfigure[$t=0.28834$]{
\includegraphics[width=0.4\textwidth]{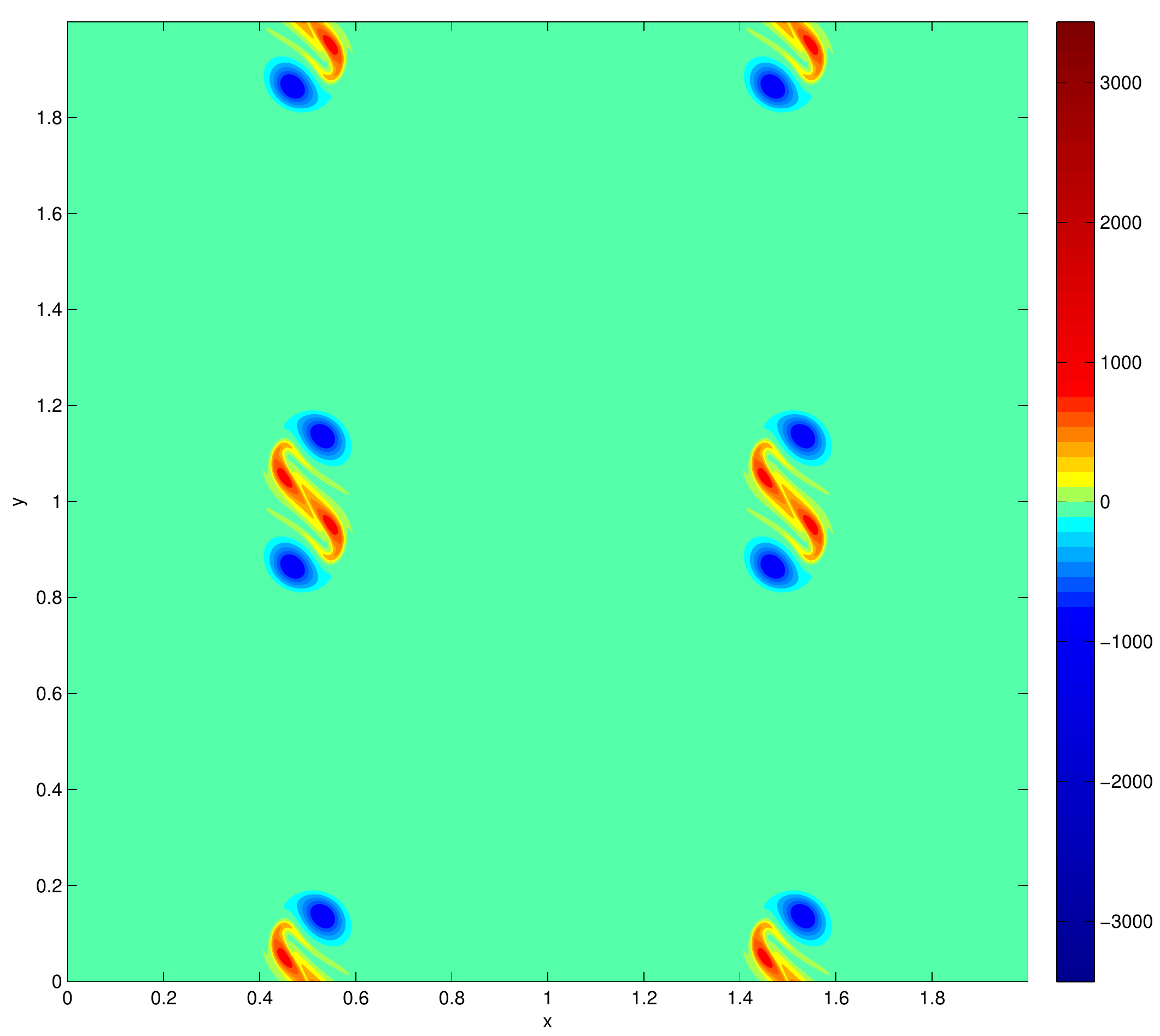}}\qquad
\subfigure[$t=0.30827$]{
\includegraphics[width=0.4\textwidth]{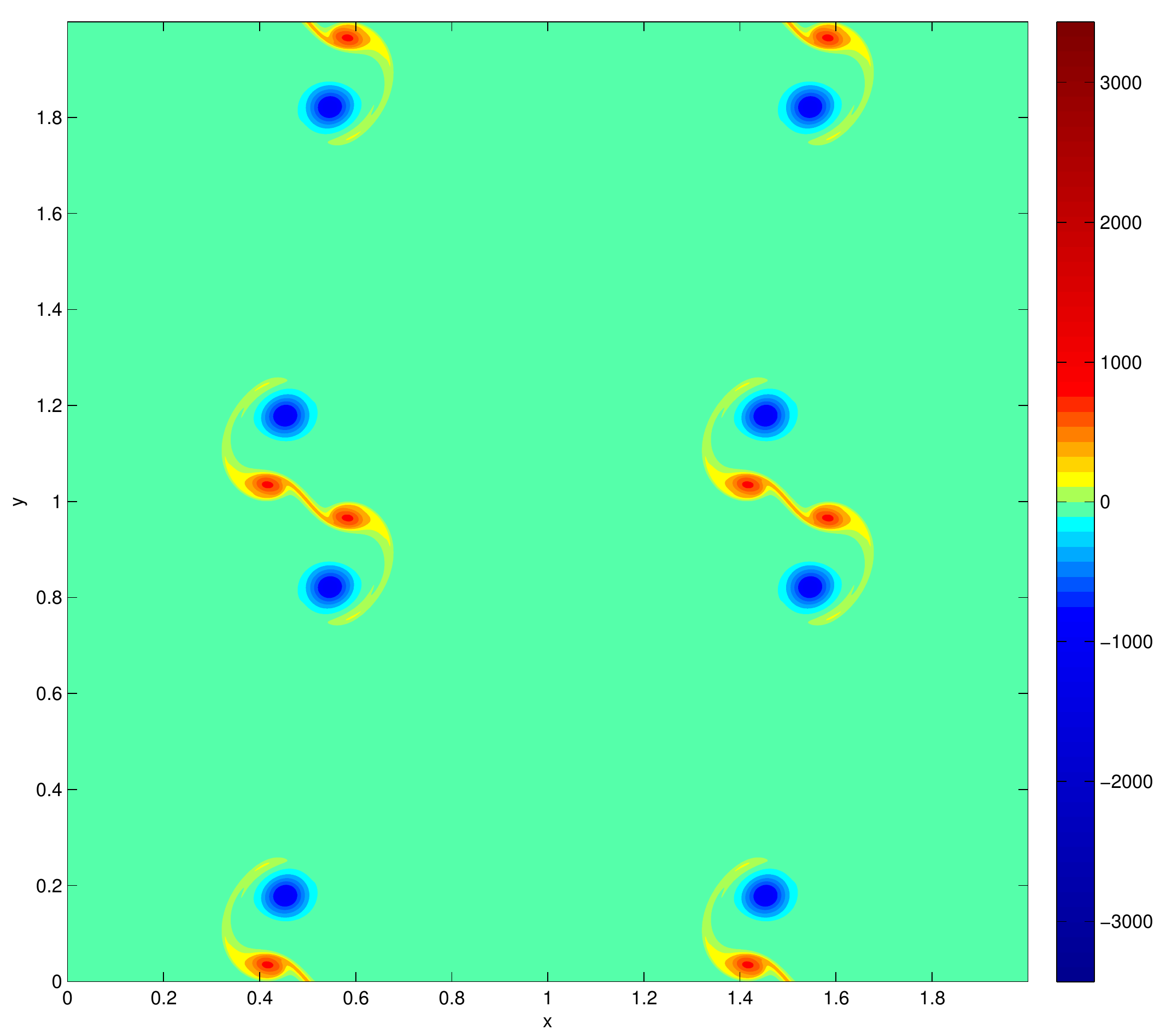}}}
\caption{Vorticity fields at the times indicated, and marked with
  solid symbols in Figures \ref{fig:kep2}(a,b,c), during the long-time
  evolution. In each Figure four copies of the periodic domain
  $\Omega$ are shown. Field (a) corresponds to the end of the
  stretching event shown in Figure \ref{fig:vor1}.}
\label{fig:vor2}
\end{figure}
\begin{figure}

\begin{center}
  \includegraphics[width=0.4\textwidth]{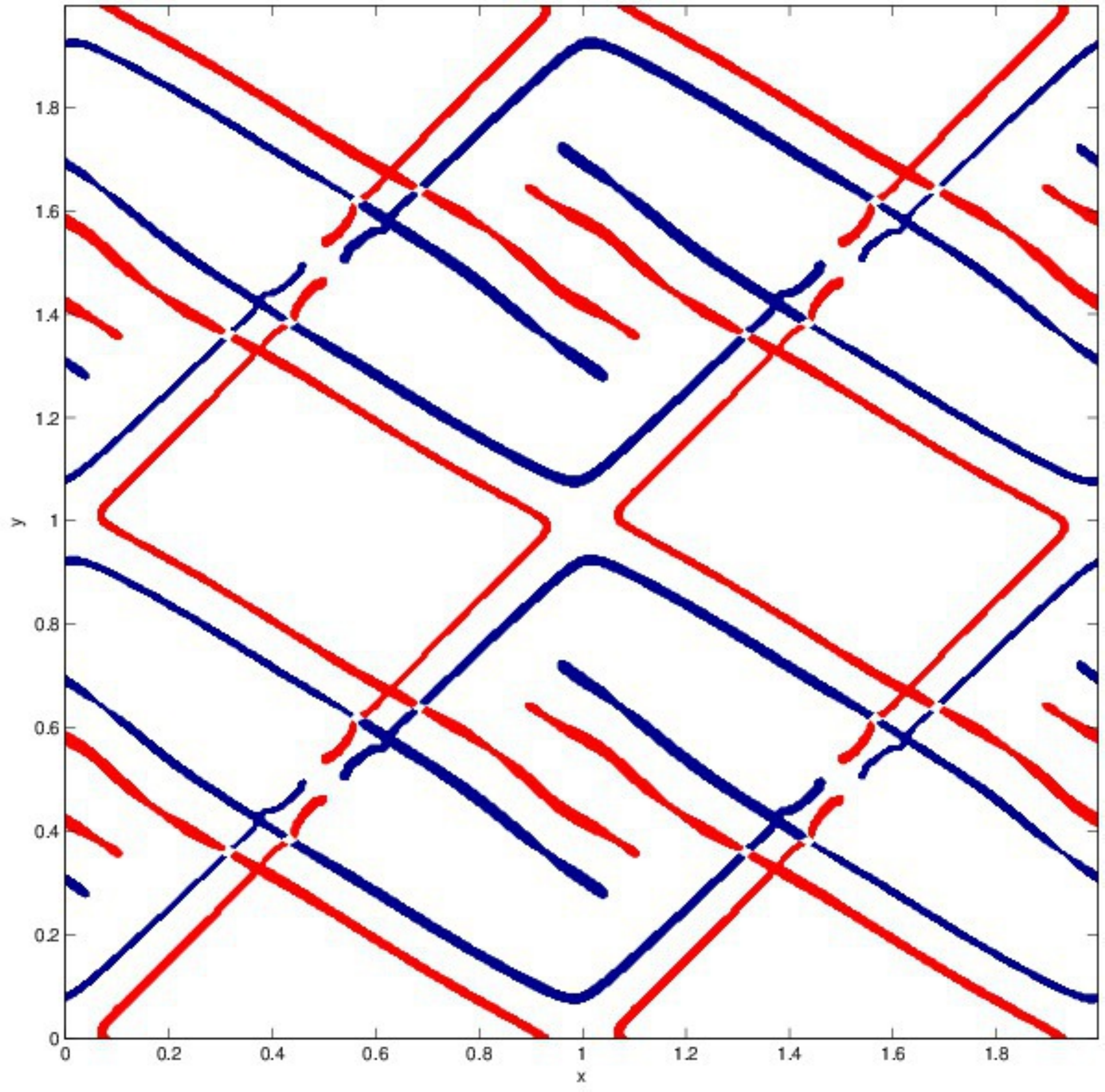}
  \caption{Traces of the vortex cores during the long-time evolution
    corresponding to the results in Figures \ref{fig:kep2} and
    \ref{fig:vor2}. At any given time, the vortex core is defined as
    the region of the domain $\Omega$ where the eigenvalues of the
    velocity gradient tensor $\bnabla\u$ are complex and have
    magnitude in the range of $90\% - 100\%$ of the maximum eigenvalue
    in the entire domain.}
\label{fig:trajectories}
\end{center}
\end{figure} 

We begin our analysis of the long-time evolution from the final state
discussed above. The history of energy $\K(t)$, enstrophy $\E(t)$ and
palinstrophy $\P(t)$ is shown in Figures \ref{fig:kep2}(a,b,c),
whereas the corresponding vorticity fields are presented in Figure
\ref{fig:vor2}. We add that, in order to discount the effect of the
vorticity dissipation during the long-time evolution, in Figure
\ref{fig:vor2} we use a different color scale than in Figures
\ref{fig:K0P0} and \ref{fig:vor1}. In addition, for clearer
presentation of the translating vortices, four copies of the periodic
domain $\Omega$ are shown in Figure \ref{fig:vor2}. In Figure
\ref{fig:trajectories} we also show the trajectories of the four main
vortices together with their periodic images (see the Figure caption
for the definition of the ``vortex cores''). Animations showing the
long-time evolution of the vorticity field and the trajectories of the
vortex cores are available on-line as supplementary material.

Following the initial stretching event discussed in Section
\ref{sec:short}, the vortices move apart as two dipoles to undergo a
``scattering event'' when they collide and then again move away after
exchanging partners (Figures \ref{fig:vor2}(b,c)). As is evident from
Figure \ref{fig:kep2}(c), the palinstrophy $\P(t)$ continues to decrease
during this event. Its otherwise steady decrease is punctuated by some
stretching events occurring later on, such as the event illustrated in
Figures \ref{fig:vor2}(d,e,f). The pattern exhibited by the trajectories
of the vortex cores shown in Figure \ref{fig:trajectories} is
reminiscent of the collision dynamics of pairs of point vortices
studied by \citeasnoun{ea88}. In this regard it should be noted that
the present problem is ``defective'', in the sense that the vortices
making up the dipoles are not identical (cf.~Figure \ref{fig:vor1}(f)).
This is a result of the asymmetry of the initial optimal configuration
$-\Delta\tpKP$ (Figure \ref{fig:vor1}(a)).

\section{Discussion and Conclusions}
\label{sec:final}

In this work we have focused on the evolution of the vorticity field
starting from the initial data $-\Delta\tpKP$ which maximizes the
instantaneous rate of palinstrophy production $d\P/dt$ under the
constraints of fixed energy $\K_0$ and palinstrophy $\P_0$. We
identified the physical mechanism leading to the growth of
palinstrophy $(\P_{\max} - \P_0)$ which in terms of the initial
enstrophy $\E_0$ is as large as allowed by the mathematically rigorous
estimate \eqref{eq:maxPt_Ayala}. Although here we presented the results
for one case only, the stretching mechanism at work at short times is
quite robust and was also observed in the short-time evolution
corresponding to the initial data $-\Delta\tpKP$ with different values
of $\K_0$ and $\P_0$. On the other hand, details of the long-time
evolution could be quite different in these different cases. For
example, for some other values of $\K_0$ and $\P_0$, during the
scattering event (cf.~Figure \ref{fig:vor2}(b,c,d)) the vortices would
spin around each other before moving apart with or without exchanging
partners, cf.~\citeasnoun{ea88}.  Classifying these different
behaviors may be an interesting problem for future research.  Another
open question is whether the secondary stretching event observed at
large times (between $t=0.25$ and $t=0.32$) in Figure
\ref{fig:kep2}(c), see also Figures \ref{fig:vor2}(d,e,f), saturates
bound \eqref{eq:dPdt_KP}. In order to answer this question, we would
need to have data characterizing $d\P/dt$ for some fixed energy $\K$
and the palinstrophy $\P$ varying over some range. Since this
stretching event is occurring at large times, there appears to be no
easy way to impose these constraints.  The long-term interest of the
present study is in providing insights about the nature of extreme
vortex events which can be useful for addressing similar questions for
the flows governed by the 3D Navier-Stokes system.

\section*{Acknowledgments}

The authors are indebted to Charles Doering and Evelyn Lunasin for
many enlightening discussions concerning the research program
mentioned in this work. This research was funded through an Early
Researcher Award (ERA) and the computational time was made available
by SHARCNET.



\bibliographystyle{jphysicsB_withTitles}

\end{document}